\documentclass[11pt, letterpaper]{article}
\pdfoutput=1
\usepackage{hyperref}
\usepackage{authblk}
\usepackage{algpseudocode}
\usepackage{algorithm}
\usepackage{amsmath}
\usepackage{amssymb}
\usepackage{amsthm}
\usepackage{color}
\usepackage{cite}
\usepackage{graphicx}
\graphicspath{{./}{Figs/}}
\usepackage{subcaption}
\usepackage{mathrsfs}
\usepackage{verbatim}
\usepackage{indentfirst}
\usepackage{appendix}

\newfont{\bbb}{msbm10 scaled 500}

\newfont{\bb}{msbm10 scaled 1100}

\newcommand{\RR}{\mbox{\bb R}}

\newcommand{\cv}{{\bf c}}
\newcommand{\dv}{{\bf d}}

\newcommand{\fv}{{\bf f}}

\newcommand{\lv}{{\bf l}}

\newcommand{\nv}{{\bf n}}

\newcommand{\sv}{{\bf s}}

\newcommand{\uv}{{\bf u}}
\newcommand{\wv}{{\bf w}}
\newcommand{\vv}{{\bf v}}
\newcommand{\xv}{{\bf x}}
\newcommand{\yv}{{\bf y}}

\newcommand{\zerov}{{\bf 0}}

\newcommand{\Cc}{{\cal C}}

\newcommand{\alphav}{\hbox{\boldmath$\alpha$}}

\newcommand{\xiv}{\hbox{\boldmath$\xi$}}

\definecolor{OXO-emph}{RGB}{153,0,0}

\newcommand\floorb[1]{\left\lfloor #1 \right\rfloor}

\title{On the organization of grid and place cells: Neural de-noising via subspace learning}\date{}\vspace{-8ex}
\author[1]{David M. Schwartz\thanks{dmschwar@email.arizona.edu}}
\author[2]{O. Ozan Koyluoglu\thanks{ozan.koyluoglu@berkeley.edu}}
\affil[1]{Dept. of Electrical and Computer Engineering, University of Arizona}
\affil[2]{Dept. of Electrical Engineering and Computer Science, University of California, Berkeley}

\begin{document}

\maketitle

\begin{abstract}
Place cells in the hippocampus are active when an animal visits a certain location (referred to as a place field) within an environment. Grid cells in the medial entorhinal cortex (MEC) respond at multiple locations, with firing fields that form a periodic and hexagonal tiling of the environment. The joint activity of grid and place cell populations, as a function of location, forms a neural code for space.
An ensemble of codes is generated by varying grid and place cell population parameters. For each code in this ensemble, codewords are generated by stimulating a network with a discrete set of locations. In this manuscript, we develop an understanding of the relationships between coding theoretic properties of these combined populations and code construction parameters. These relationships are revisited by measuring the performances of biologically realizable algorithms implemented by networks of place and grid cell populations, as well as constraint neurons, which perform de-noising operations.
Objectives of this work include the investigation of coding theoretic limitations of the mammalian neural code for location and how communication between grid and place cell networks may improve the accuracy of each population's representation. Simulations demonstrate that de-noising mechanisms analyzed here can significantly improve fidelity of this neural representation of space. Further, patterns observed in connectivity of each population of simulated cells suggest that inter-hippocampal-medial-entorhinal-cortical connectivity decreases downward along the dorsoventral axis.
\end{abstract}

\section{Introduction}
\label{sec:intro}
Place cells are a classs of spatially modulated neuron with an approximately bivariate Gaussian tuning curve centered on a particular location in the environment, and have been identified in the hippocampus \cite{OKeefe1971, o1976place, EkstromKahanaCaplanEtAl2003}. Grid cells are spatially modulated neurons with firing fields that form a periodic and hexagonal tiling of the environment, and are found in the Entorhinal Cortex (EC) of rats, mice, bats, and humans \cite{Hafting2005, FyhnHaftingWitterEtAl2008, YartsevWitterUlanovsky2011, doeller2010evidence, JacobsWeidemannMillerEtAl2013}. Grid cells are clustered in discrete modules wherein cells share grid scale \cite{StensolaStensolaSolstadEtAl2012}. Anatomically, both cell types share a dorsoventral organization, with cells possessing wider receptive fields distributed towards the ventral end \cite{Strange2014, StensolaStensolaSolstadEtAl2012}. It is known that the rat grid cell network requires communication from the hippocampus to maintain grid-like activity \cite{BonnevieDunnFyhnEtAl2013}, and that a significant improvement in accuracy of the rodent place cell representation is tightly correlated with the emergence of the grid cell network \cite{MuessigHauserWillsEtAl2015}. However, the mechanisms by which these networks communicate and how each may bolster the other's accuracy are unknown. Objectives of this work include the investigation of coding theoretic limitations of the mammalian neural code for location and how communication between grid and place cell networks may improve the accuracy of each population's representation.\par
Associative memories are a class of biologically implementable content addressable memory consisting of networks of neurons, a learning rule, and in some instances, a separate recall process \cite{Hopfield1982, Amit1989}. This means that they can be exploited to stabilize the states of their constituent neurons to match a previously memorized network state if enough of the network already lies in this state. The information capacity of the simplest of these constructions is quite limited: $\frac{n}{2\log{n}}$ bits, for a network of $n$ binary neurons \cite{McEliecePosnerRodemichEtAl1987} . However, recent advances by Salavati et al. take advantage of sparse neural coding and non-binary neurons to design an associative memory with information storage capacity exponential in the number of neurons \cite{SalavatiKumarShokrollahi2014}. Sparse connectivity confers the memory network with other performance improvements: infrequent spiking implies reduced energy costs and faster convergence to a stable state.\par
In communications, this principle is leveraged by low density parity check codes (LDPC), a class of linear block code whose power (in coding and decoding complexity) depends on sparsity of the code's parity check matrix. Commonly, de-noising a LDPC code involves iteratively passing messages along edges of a bipartite graph consisting of a collection of nodes that stores and updates an estimate of the originally transmitted word connected to a collection of nodes that computes the code's parity check equations \cite{Chen2002,declercq2007decoding}. Recent developments in the intersection of coding theory and machine learning demonstrate that neural networks can learn an approximation of a LDPC code's parity structure, and by executing message passing algorithms recover memorized patterns in the presence of noise \cite{SalavatiKumarShokrollahi2014}.\par
Grid cell population activity forms a dense modulo code, in which (in the absence of noise) information about position at any grid scale may be present in the activity of each module \cite{FieteBurakBrookings2008}. Place cell activity forms a relatively sparse code (for enough cells, and a sufficiently large environment), thus combining populations of grid and place cells realizes codes that are sparser than the grid cell component of the code.\par
Nature provides myriad circumstances in which many neural computations (e.g. object recognition, acoustic source localization, and self localization) must be executed robustly in the presence of neural noise if the organism is to survive. We propose a de-noising mechanism for populations of grid and place cells, in the form of the associative memories described in \cite{SalavatiKumarShokrollahi2014}, \cite{KarbasiSalavatiShokrollahiEtAl2014}, and \cite{KarbasiSalavatiShokrollahi2013}, which takes advantage of coding theoretic properties of these populations to eliminate the negative impacts of noise. In particular, we propose a de-noising algorithm that relies on the biological organization of grid cells into discrete modules, and observe that after learning, average connectivity between place cells and grid modules decreases with increasing place cell size for each module.\par
Redundancy in receptive field (RF) population codes is known to confer improvements in decoding accuracy when a small tolerance to error is introduced (expressed in this case, in the stimulus space to which we decode) \cite{Curto2012}. We believe we are the first to investigate coding theoretic impacts of redundancy in grid cell populations. We investigate the impact of this redundancy on decoding accuracy by comparing de-noising and decoding performance across codes of varying redundancies. We demonstrate that a maximum likelihood (ML) estimator reliably decodes position from population activity with small position estimation error in the presence of bounded noise. Overall, our work shows that the biological organization of grid cells into modules may be necessary for optimal estimation of self location.\par
This paper is organized sectionally. In section \ref{sec:intro} we introduce a few key concepts and present the main results. Section \ref{sec:theoreticalFramework} introduces the theoretical framework upon which our model is built, describing the code construction, de-noising network, learning algorithms, and de-noising algorithms in sections \ref{subsec:code}, \ref{subsec:deNoisingNet}, \ref{subsec:subspaceLearning}, and \ref{subsec:denoisingAndDecoding}, respectively. Section \ref{subsec:codingResultsSection} presents results of all coding theoretic analysis and experimentation. Section \ref{subsec:learningResults} annotates results of the aforementioned learning algorithms. Section \ref{subsec:denoisingAndDecodingResults} describes outcomes of performance tests of the de-noising algorithms. Section \ref{sec:discussion}, consists of discussion of these results, their implications, limitations, and a physiologically testable hypothesis they inform.\par

\section{Theoretical framework}\label{sec:theoreticalFramework}
\subsection{A hybrid code}\label{subsec:code}

We consider a population of place and grid cells, a total of $N$ neurons. There are $M$ grid cell modules, each module, $m$, containing $J_m$ neurons, and $P$ place cells. Throughout this manuscript, we use $J$ to refer to the number of grid cells in module $1$, which - if grid cells are allocated to modules non-uniformly - is not equal to each other module's $J_i$. The firing rate of each grid cell is denoted as $g_{m,j}$, for $m\in\{1,\cdots,M\}$ and $j\in\{1,\cdots,J_m\}$. Place cells' firing rates are denoted as $p_{i}$, for $i\in \{1,\dots,P\}$. The activity of this population, as a function of location $\ell$, is represented by
\begin{equation*}
\xv_i(\ell)=\begin{cases}
g_{m,j}(\ell), & i = \sum\limits_{k = 1}^{m-1} J_k + j, i \leq \sum\limits_{m=1}^{M} J_m \\
p_{i-MJ}(\ell), & i>\sum\limits_{m=1}^M J_m
\end{cases}
\end{equation*}
where the location dependent mean firing rates of the grid cells, $g_{m,j}(\ell)$, are given by the following two-dimensional distributions resembling von Mises density functions,
\begin{align}
g_{m,j}(\sv)= \frac{f_{\text{max}}}{Z}\text{exp}\left[  \sum\limits_{k = 1}^3 \text{cos}\left(\frac{4}{\lambda_{m}\sqrt{3}}\uv(\theta_k - \theta_{m,j}\right)\cdot(\sv -\cv_{m,j}) + \frac{3}{2} ) -1\right],
\end{align}
where $\uv(\theta_k - \theta_{m,j})$ is a unit vector in the direction of $\theta_k  - \theta_{m,j}$, $\sv \in [0,L]\times [0,L]$ is the position stimulus, $\cv_{m,j}$, $\theta_{m,j}$, and $\lambda_{m}$ are the grid cell's spatial phase offset, orientation offset, and scaling ratio. Grid cell orientations were taken to be ideal values about which the measurements presented in \cite{StensolaStensolaSolstadEtAl2012} appear to fluctuate. More precisely, we choose $\theta_{k} \in\lbrace-60^{\circ}, 0^{\circ}, 60^{\circ}\rbrace$. A scaling ratio of $\lambda$ defines the scale of module $m$ as $\lambda_m=\lambda_1(\lambda)^{m-1}$.  $Z$ is a normalizing constant ($\approx 2.857399$), and $f_{\text{max}}$ is the grid cell's maximum firing rate.
In two dimensions, place cells have bivariate Gaussian tuning curves, with mean $\xiv \in [0,L] \times [0,L]$, correlation, $\rho \in [-\frac{1}{2}, \frac{1}{2}]$ (chosen uniformly randomly), and covariance
$\bigl(\begin{smallmatrix}
\sigma_1^2&\rho\sigma_1 \sigma_2 \\ \rho\sigma_1 \sigma_2&\sigma_2^2
\end{smallmatrix} \bigr)$,
where $\sigma_1$ and $\sigma_2$ are chosen independently and uniformly randomly from $[0.9\lambda_1, 1.1\lambda_M]$. We require that $\sigma_1$ and $\sigma_2$ depend on $\lambda_1$ so that both grid and place cell receptive fields lie in similar spatial scales.\par
\begin{figure}[H]
\centering
\includegraphics[width=0.8\textwidth]{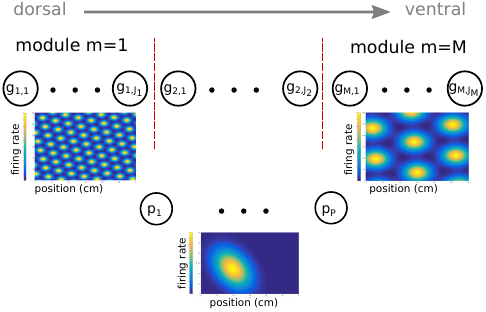}
\label{fig:2d_hybrid_code}
\caption{Concatenation of activities of grid and place cells (shown with typical idealized model receptive fields) to form the hybrid code}
\end{figure}
\paragraph{A hybrid codebook:}
$C$ codewords, of length $N = P + \sum\limits_{m = 1}^{M} J_m$, are generated by choosing locations, $\lv$,  from the vertices of a square lattice imposed on the plane, with unit area equal to $(\Delta L)^2$, and total area equal to $L^2$. $\Cc$ is assembled by placing these codewords in its rows, and represents the states of the grid and place cells when stimulated with these positions.

\subsection{De-noising network}\label{subsec:deNoisingNet}
Two high-capacity associative memory designs are considered to test the hybrid code's resilience to noise. In each case, the memory network is a bipartite graph consisting of $N$ pattern neurons (i.e., grid and place cells) and $n_i$ constraint neurons. In the un-clustered design, all constraint neurons are connected to a random set of pattern neurons. In the clustered configuration, the constraint neurons  were split into $M$ distinct clusters of $n$ constraint neurons per cluster, with each cluster connected to a distinct grid module.  Each cluster's constraint neurons were connected randomly to pattern neurons, chosen from a set consisting of every grid cell in the corresponding module, and every place cell.\par
\begin{figure}[H]
\centering

	\begin{subfigure}[b]{0.75\textwidth}
	\centering
	\subcaption{}
	\includegraphics[scale=1.75]{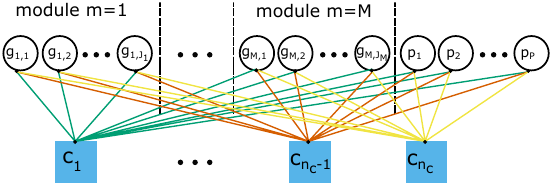}
	\label{fig:unclusteredAssMemFig}
	\end{subfigure}\
	\begin{subfigure}[b]{0.75\textwidth}
	\subcaption{}
	\centering
	\includegraphics[scale=1.75]{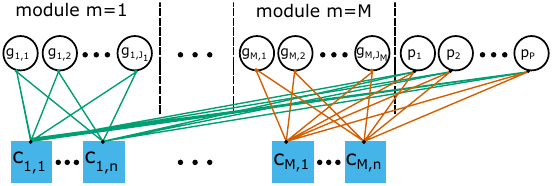}
	\label{fig:clusteredAssMemFig}
	\end{subfigure}
\caption{(a) Structure of an un-clustered de-noising network - considered as a baseline for comparison to the neurophysiologically inspired systematic clustering scheme (b) Structure of a systematically clustered de-noising network in which clusters of constraint neurons connect to all place cells but only to the corresponding module of grid cells.}
\end{figure}
We also consider a foil to this systematic clustering scheme organized by grid modules: Grid and place cells are randomly assigned to clusters. Figures \ref{fig:unclusteredAssMemFig} and \ref{fig:clusteredAssMemFig} depict the general connectivity structure of the un-clustered and clustered designs, respectively. In both the clustered and un-clustered configurations, a neurally plausible modified version of Oja's subspace learning rule was applied to learn the code, i.e., a sparse connectivity matrix is found such that the weights of connections from constraint neurons to pattern neurons lie orthogonal to the code space (i.e., the space spanned by $\Cc$) \cite{OjaKohonen1988}. This way, constraint neuron connectivity converges to the parity structure of the code and may be utilized in de-noising operations.\par

\subsection{Code construction via subspace learning}\label{subsec:subspaceLearning}
Before we can use the de-noising system to correct corrupted codewords, it must learn (i.e., adapt its weights for) the hybrid code. This process is complete when the constraint neurons may be read to determine if the states of the pattern neurons map to a valid codeword. Formally, this amounts to finding a connectivity matrix, $W$ ($W_{i,j}$ is the synaptic weight between constraint neuron $i$ and pattern neuron $j$), whose rows are approximately perpendicular to the code space. A procedure to procure such a matrix is outlined in \cite{OjaKohonen1988}, and improved in \cite{SalavatiKumarShokrollahi2014}. Note here that this learning process is not a model for the development of either grid or place cells' apparent receptive fields nor their remapping, as in \cite{Monaco2011}. These algorithms begin with a random set of vectors, and for each, seeks a nearby vector orthogonal to $\Cc$ (i.e., a vector onto which each element of $\Cc$ has minimal projection). We implement this in Algorithm \ref{alg:algorithm_1} (a derivation of this algorithm can be found in the appendix). In the clustered design, Algorithm \ref{alg:algorithm_1} is applied to each cluster's local connectivity matrix. Note that here, all arithmetic on the synaptic weights, $W_{i,j}$ is performed in $\RR$, while arithmetic on states of neurons (i.e., their firing rates), is quantized to the nearest integer in $[0, Q-1]$. The maximum firing rate, $f_{\text{max}} = Q-1$, is identical for all neurons. With each update, $\wv \leftarrow \wv - \alpha_t (y(\xv - \frac{y \wv}{\Vert \wv \Vert^2}) + \eta \Gamma(\wv, \theta))$, where $\theta$ is a sparsity threshold, $\eta$ is a penalty coefficient, $y = \xv^T\wv$ is the scalar projection of $\xv$ onto $\wv$, and $\alpha_t$ is the learning rate at iteration $t$. $\Gamma$ is a sparsity enforcing function, approximating the gradient of a penalty function, $g(\wv ) = \sum\limits_{k = 1}^{m}\tanh(\sigma {\wv_k}^2)$, which, for appropriate choices of $\sigma$, penalizes non-sparse solutions early in the learning procedure \cite{SalavatiKumarShokrollahi2014}. \par
\begin{algorithm}[H]
\caption{Neural Learning}
\label{alg:algorithm_1}
	\begin{algorithmic}[1]
	\Require {set of $C$ patterns, $\Cc$, stopping point, $\epsilon$}
	\Ensure {learned weights matrix, $W$}
			\For{rows, $\wv$, of $W$}
				\For{$t \in \lbrace 1,...,T_{\text{max}} \rbrace$}
					\State $\alpha_t \gets \text{max}\lbrace {\frac{50\cdot \alpha_0}{50+\log_{10}(t)} , 0.005} \rbrace$
					\State $\theta_t \gets \frac{\theta_0}{t}$
					\For{$\cv\in \Cc$}
						\If{$\Vert \cv \Vert > \epsilon$}
							\State $\alpha_t \gets \frac{\alpha_0}{\Vert \cv \Vert ^2}$
						\EndIf
						\State $\wv \gets \text{Dale}(\text{update}(\bold{\cv}, \wv, \alpha_t, \theta_t, \eta))$
					\EndFor
					\If{$\Vert \underbar{\Cc}\wv' \Vert < \epsilon$}
						\State break
					\EndIf
					\State $t \gets t + 1$
				\EndFor
			\For{components, $w_i$ of $\wv$}
				\If{$\vert w_i \vert \leq \epsilon$}
					\State$\wv_i \gets 0$
				\EndIf
			\EndFor
			\EndFor
	\end{algorithmic}
\end{algorithm}
As in \cite{SalavatiKumarShokrollahi2014}, to speed up learning, we approximate $\Gamma = \nabla g$ with \begin{displaymath}
	\Gamma(w_t, \theta_t) = \left\{
	\begin{array}{lr}
		w_t & : \vert w_t \vert \leq \theta_t \\
		0 & : \text{otherwise}
	\end{array}
	\right.
	\end{displaymath}
This update rule is a an improved approximation to Oja's Hebbian learning algorithm \cite{OjaKohonen1988}, with advantages in both biological plausibility and computational complexity. For connections of fixed type (i.e., inhibitory vs. excitatory), Oja's rule alone is biologically dubious without the inclusion of many constraint neurons to manage this change in type. Dale's Principle states that real synaptic connections change type rarely, if ever \cite{Eccles1976}. In accordance with this principle, our update rule does not allow weights to change sign. This is accomplished after the updated weights are determined: If the sign has changed after applying the update, set the new weight to a value just above (resp. just below) zero if the previous weight was positive (resp. negative). Thus, when learning is complete, these weights will be small in magnitude and are thresholded to zero.\par
In Algorithm \ref{alg:algorithm_1}, line $12$ terminates learning of the current weight, $\wv$, if the sum of the projections of $\wv$ on each pattern is no more than $\epsilon$ away from zero, that is, if the current weight vector is approximately orthogonal to the code space. Lines $17$-$19$ perform a thresholding operation that maps to zero any weight sufficiently small in magnitude. This is primarily to suppress numerical errors and promote consistency, as in Line $11$, we use $\epsilon$ as a small positive constant. Note that since the weights processed on each iteration are independent of those in other iterations, this algorithm can be readily parallelized so that each constraint neuron learns its weights simultaneously.\par

\subsection{De-noising and decoding}\label{subsec:denoisingAndDecoding}
We implemented a Bit Flipping style neural de-noising process, which we applied to both the clustered and un-clustered de-noising networks. For all configurations (clustered and un-clustered, and for a fixed maximum number of de-noising iterations, the bit flipping algorithm has perform no worse than winner-take-all. Moreover, since it requires only the additional implementation parallel thresholding operations for each pattern neuron, a biological realization of their inclusions is no less plausible. The goal of this algorithm is to recover the correct activity pattern, $\xv$, which has been corrupted by noise, and as such, is currently (and errantly) represented by a noisy version, $\xv_n = \xv + \nv$, where $\nv$ is this noise pattern. Since each weight vector is nearly perpendicular to every pattern, for a matrix of weights, $W$ , $\xv_n W'$ reveals inconsistencies in $\xv_n$, which the de-noising algorithm seeks to correct in the feedback stage \footnote{To see this, consider that $\xv_n W' = (\xv + \nv)W' = \xv W' + \nv W' \approx 0 + \nv W'$}. In de-noising, feedback weights from constraint neurons to pattern neurons are taken to be equal to the corresponding feed-forward weight (i.e., synaptic connectivity is symmetric). The clustered de-noising process begins with Algorithm \ref{alg:sequential_denoising}, in which each cluster attempts to detect errant pattern neurons. If no errors are detected, the process is complete. Otherwise, algorithm \ref{alg:modular_recall} is invoked for each cluster that detected errant neurons. This and other de-noising processes are discussed in greater detail in \cite{KarbasiSalavatiShokrollahi2013} and \cite{SalavatiKumarShokrollahi2014}. Note that this de-noising mechanism differs from error correction methods presented in \cite{FieteBurakBrookings2008} and \cite{StemmlerMathisHerz2015} in that information contributed by place cells only reaches grid cells through constraint neurons, and place information contributed by grid cells at module $i$ only reaches other modules through constraint neurons if connectivity allows.\par
In order to quantify the information content of the population, we estimated the location encoded by the population using a maximum likelihood decoder in 4 different schemes. Joint hybrid decoding utilizes information from all cells. Grid (resp. place) only decoding utilizes information from only grid (resp. place) cells. Grid decoding conditioned on place response performs decoding using only information provided by the grid cells, however, the only candidate locations considered for the estimate are those that are not impossible given the place cell activity.\par
\begin{algorithm}[H]
\caption{Modular Recall}
\label{alg:modular_recall}
	\begin{algorithmic}[1]
		\Require{local weights for this cluster, $W$, maximum number of iterations, $T_{\text{max}}$, noisy subpattern, $\xv$, feedback threshold, $\phi$}
		\Ensure{denoised subpattern, $\dv$}
		\State $\dv \gets \xv$
		\While{$t < T_{\text{max}}$}
			\State $\yv \gets \xv W' $
			\If{$\Vert \yv \Vert < \epsilon$}
				\State break;
			\EndIf

			\State $\fv \gets \frac{\vert \yv\prime \vert \cdot \vert W \vert} {\sum\limits_{i = 1}^m \vert W \vert }$

			\For{$\text{each pattern neuron,} j$}
				\If{$\fv_{j} \geq \phi$}
					$\fv_{j} = \text{sign}(\xv_{j})	$
				\Else{}
					$\fv_{j} = 0$
				\EndIf
			\EndFor

			$\dv \gets \dv - \fv$
		\EndWhile
	\end{algorithmic}
\end{algorithm}

\begin{algorithm}[H]
\caption{Sequential de-noising}
\label{alg:sequential_denoising}
	\begin{algorithmic}[1]
		\Require{local weights, $W_i$, for each cluster, $i \in \lbrace 1,..., M \rbrace$, noisy pattern, $\xv_{n}$, stopping threshold, $\epsilon$}
		\Ensure{denoised pattern, $\xv_{d}$}
		\State $\xv_{d} \gets \xv_{n}$
		\While{$t < T_{\text{max}}$ and a cluster has an unsatisfied constraint}

			\For{each cluster, $i \in \lbrace 1,..., M \rbrace$}
				\State $\xv \gets \text{subpattern corresponding to cluster } i$
				\State $\dv \gets \text{Modular}\_\text{Recall}(\xv, W_i)$
				\If{$\vert \dv W_i \vert \leq \epsilon$}
					\State $\xv_{d}(\text{cluster i's subpattern indices}) \gets \dv$
				\EndIf
			\EndFor
		\State $t \gets t + 1$
		\EndWhile
	\end{algorithmic}
\end{algorithm}

\section{Results}
\label{sec:Results}
\subsection{Coding theoretic results}
\label{subsec:codingResultsSection}
We now endeavor to disentangle the impacts of grid and place cell parameter choices on coding theoretically relevant dependent variables, and the way that these are linked. We begin our investigation of coding theoretic properties of the hybrid code by defining two measures of redundancy for grid cells: $\mu_p$ and $\mu_o$. Define $\mu_p$, a hybrid code's spatial phase multiplicity, as the number of grid cells with the same phase in the same module. In \cite{Wennberg2015}, it is revealed that there may be a highly non-uniform distribution of phases among grid cells. Considering replication of grid cells (i.e. modules consisting of multiple grid cells of the same phase) allows us to investigate coding theoretic repercussions of this phenomenon. We also set $\mu_o = \floorb{\frac{J}{3}}$, the code's orientation multiplicity (i.e., minimum number of replications of a orientation offset in the first module). These replications are the main sources of grid cell redundancy under consideration. Inspired by \cite{MosheiffAgmonMorielEtAl2017}, for each of these schemes, we consider two distributions of grid cells to modules: uniform and non-uniform. Mosheiff et. al find in \cite{MosheiffAgmonMorielEtAl2017} that choosing $J_m \propto \frac{1}{\lambda_m}$ produces a more efficient representation of space. When modelling the non-uniform allocation of grid cells to modules, we chose $J_m = \floorb{\frac{J}{\lambda_1(\lambda^{m-1})}}$ as the scale of module $m$ is defined as $\lambda_m=\lambda_1(\lambda)^{m-1}$. Neural recordings show that the smallest scale is $\lambda_1\approx 40$cm (the value used here) \cite{StensolaStensolaSolstadEtAl2012}.\par
We construct a codebook matrix, $\underbar{\Cc}$, by placing elements of $\Cc$ in its rows. We computed normalized rank of the code, $R = \frac{\text{rank}(\underbar{\Cc})}{N} \in [0,1]$ as a function of the grid scaling ratio. Normalized rank is an indicator of a code's density, expressed as the fraction of possible dimensions of the code space occupied by a particular code. $R$ is an important feature to consider since a code's dimensionality determines the dimensionality of it's null space, the object that is learned by the de-noising network. As discussed in \cite{SalavatiKumarShokrollahi2014}, if we suppose that $\Cc \subset \RR^{n}$, and dim$(\Cc) = k < n$, then there are $n-k$ mutually orthogonal vectors that are also orthogonal to our code space (e.g. any basis for the null space of the code), each representing one valid constraint equation. Thus $R$ provides a fundamental limit on the number of unique effective constraint nodes we the de-noising network may learn.\par
The grid cell code is known to be dense \cite{FieteBurakBrookings2008}. This is especially pronounced when all orientations and phases are chosen randomly (uniformly from $[0, 2\pi]$ and $[0,L]\times[0,L]$ respectively), where for all choices of other parameters, the hybrid code achieves full rank at low rate. That is to say that the experimentally observed properties of the grid cell code described in \cite{StensolaStensolaSolstadEtAl2012} produce a measurable decrease in $R$ compared to typical ranks observed when orientations and phases are chosen randomly.\par
\begin{figure}[H]
\centering
\includegraphics[width=15cm,height=5cm]{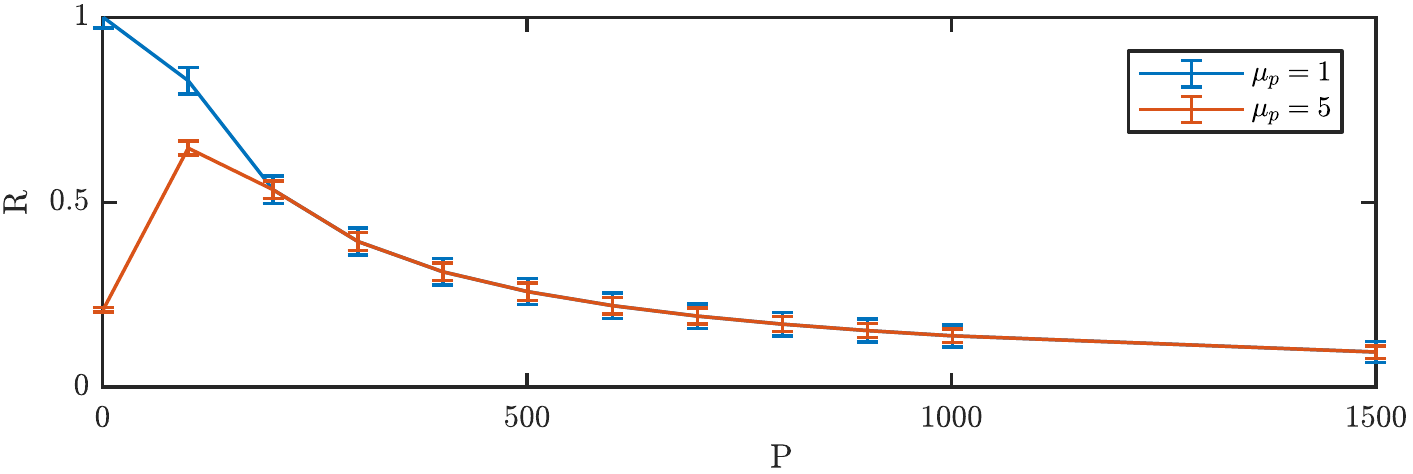}
\caption{Code rank ($R$) vs. number of place cells, $P$, for a uniform allocation of grid cells. Increasing $\mu_p$ produces a code with low rank until sufficiently many place cells are included in the code that additional place cells contribute only redundant location information. Here (and in any other plot containing them) error bars show standard error of the mean.}
\label{fig:rankVsP}
\end{figure}
\begin{figure}[H]
\centering
\includegraphics[width=8cm,height=7.5cm]{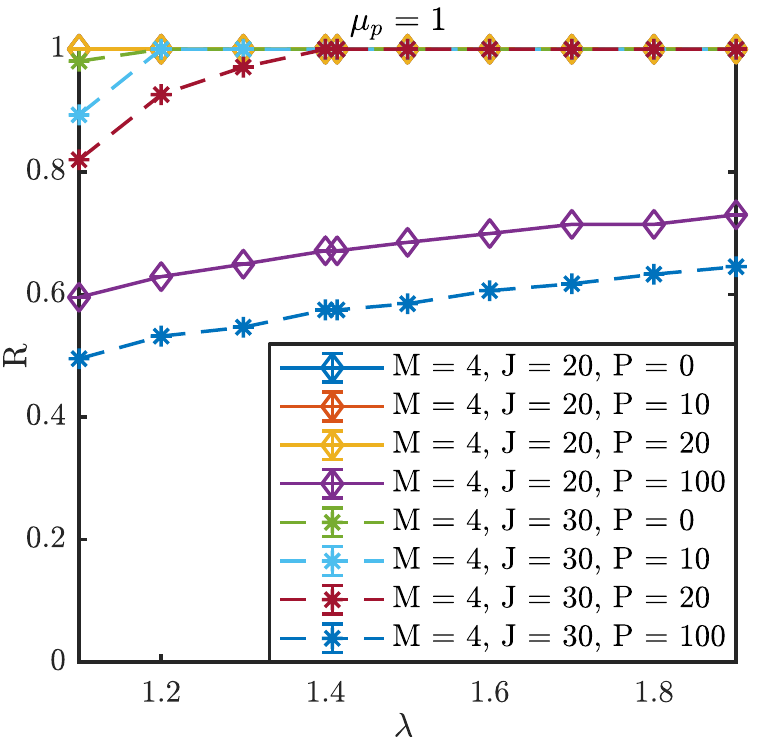}
\includegraphics[width=8cm,height=7.5cm]{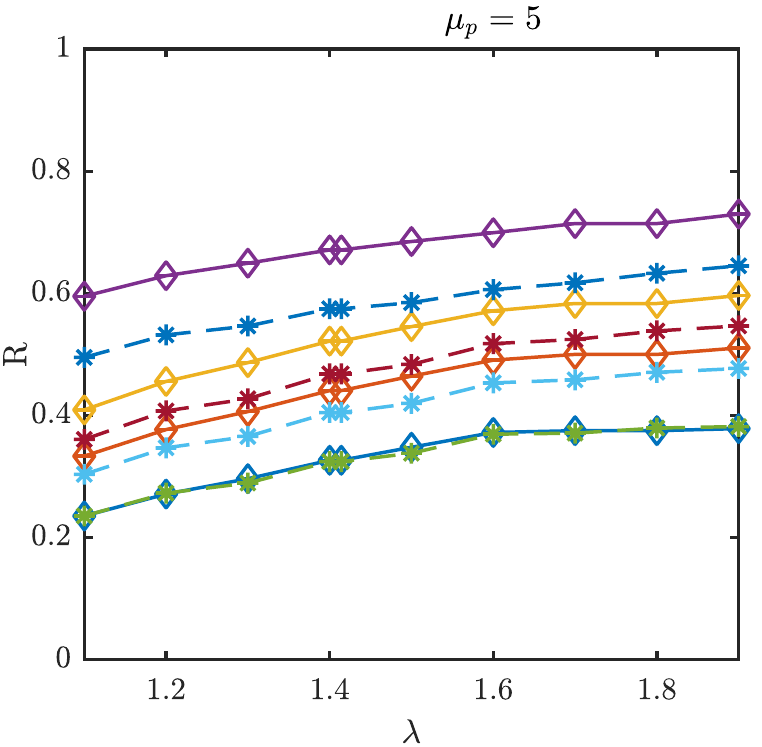}
\caption{Code rank ($R$) vs. scaling ratio ($\lambda$) for a non-uniform allocation of grid cells to modules for $r = 1$, for $\mu_o \in \lbrace 6, 10\rbrace$. $R$ is insensitive to changes in $\lambda$ but strongly affected by changes in $N$ and $\mu_p$. Nearly identical trends are also observed when grid cells are allocated to modules uniformly. Interestingly, when phase redundancy ($\mu_p$) is minimized, minimal rank is achieved when $\mu_o$ is maximized. However when $\mu_p$ is maximal (i.e., $\mu_p = 5$), changes in $\mu_o$ have little effect on $R$.}
 \label{fig:rankVsLambda}
\end{figure}

When $\mu_p = 1$, a hybrid code with no place cells achieves the largest normalized rank. Since place cells communicate redundant information, their inclusion also reduces rank, which is precisely the trend observed in Figure \ref{fig:rankVsP}. However, this appears to reverse when $\mu_p > 1$, for a sufficiently small number of participating place cells. This occurs because rendering grid cells redundant lowers the rank of the grid-only hybrid code. Consequently, including place cells increases rank, until the information contributed by the place cells reaches its maximum, at which point the inclusion of additional place cells only lowers rank. In contrast to the uniformly allocated grid cells, non-uniformity shifts the rank curve up, though the effect is less pronounced as $\mu_p$ increases. In all cases, for a fixed rate, code rank has little dependence on $\lambda$. Figure \ref{fig:rankVsLambda} illustrates the influence of $\mu_o$ on rank. Increasing $\mu_o$ decreases rank, even for less dramatic changes than those shown. This is because increases in $\mu_o$ correspond to increases in $N$, to which $R$ is inversely proportional. It is this common demoninator ($N$) that links the dependence on population size of both rank and rate ($R$ and $r$ respectively). Interestingly, when phase redundancy ($\mu_p$) is minimized, minimal rank is achieved when $\mu_o$ is maximized. However when $\mu_p$ is maximal (i.e., $\mu_p = 5$), changes in $\mu_o$ have little effect on $R$. Error bars (measuring SEM) are included due to the stochastic nature of instantiating certain hybrid code parameters (e.g. $\xiv$, which is always chosen uniformly randomly from the set of quantized locations).\par
\begin{figure}[H]
\centering
\includegraphics[width=15cm, height=7.5cm]{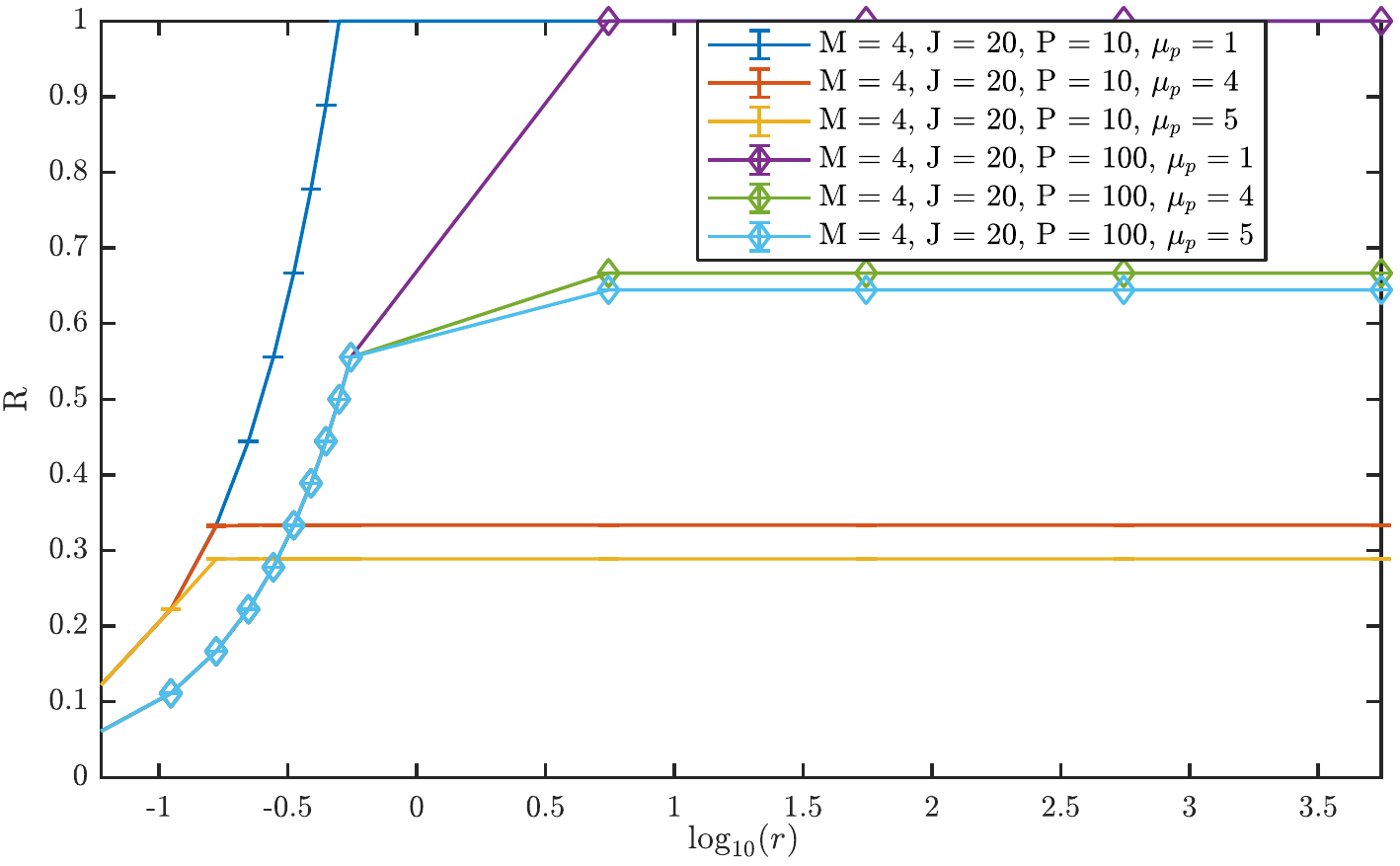}
\caption{Code rank ($R$) vs. log of code rate ($\log_{10}(r)$) for grid cell organization choices consistent with observations in \cite{StensolaStensolaSolstadEtAl2012} with $M = 4$, $J = 20$, $P \in \lbrace 10, 100\rbrace$, $\mu_p \in \lbrace 1,4,5\rbrace$, and $\mu_o = 6$. Without grid cell phase redundancy, rank saturates for very small rates. In contrast, when phase redundancy is imposed on the grid cell population, low ranks are achievable at a wide range of rates.}
\label{fig:rankVsRate}
\end{figure}

We also computed rank, $R$, as a function of code rate, $r =\frac{C}{N}$ (number of locations represented per neuron), measures spatial resolution, i.e., for a fixed $L$, a higher code rate, $r$, is obtained by lowering $\Delta L$ or by decreasing $N$. That is, $r$ measures efficiency of an encoding of position when the only cost considered is the number of neurons required to encode this position. We expect that $R$ is a non-decreasing function of $r$ as each is inversely proportional to $N$. When phases are chosen randomly, low rank is difficult to obtain at all but the smallest of code rates tested ($r \in [0,1]$ and $\mu_p > 1$ may result in low ranks if enough place cells are included). In contrast, Figure \ref{fig:rankVsRate} shows that codes spanning the spectrum of normalized ranks may be instantiated over a wide range of rates with appropriate choice of parameters. Further, this indicates that redundancy reduces dimensionality so low ranks are achievable even at rates much greater than biologically relevant. Later, it will be shown that this low dimensionality is important in constructing sparse and readily de-noisable representations of space. Figure \ref{fig:rankVsRate} demonstrates that without the redundancy introduced by increasing $\mu_p > 1$, a hybrid code that encodes in $90$ neurons more than $90$ locations in a $9$ $m^2$ environment has full rank. However when $\mu_p > 1$, there is a stark drop in the maximum rank achieved. As shown, when $\mu_p > 1$, one may encode orders of magnitude more locations while maintaining low dimensionality, with relatively fewer cells. This trend is observed in each configuration shown when sufficient replication of grid phase is ensured. Therefore both dense and sparse hybrid codes may be instantiated by proper choice of population size and redundancy parameters.\par
\begin{figure}[H]
\centering
\includegraphics[width=15cm,height=7.5cm]{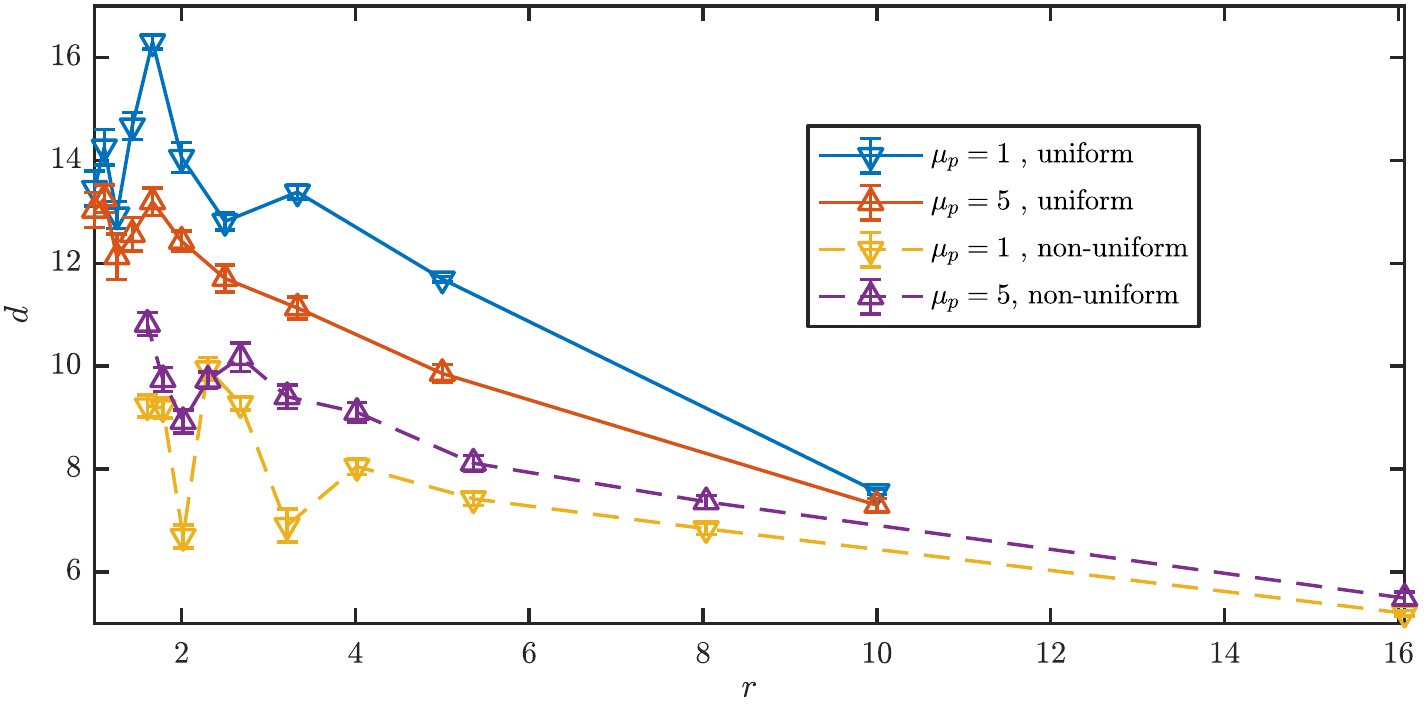}
\caption{Minimum distance ($d$) vs. $r$, for grid cell organization consistent with observations in \cite{StensolaStensolaSolstadEtAl2012}, with $M = 4$, $J = 20$, and $P = 10$. In all cases, $d$ decreases with increases in $r$ and a non-uniform allocation of grid cells across modules diminishes minimum distance.}
\label{fig:d_min_vs_rate}
\end{figure}

A code's resilience to neural noise can be assessed by the minimum  pairwise (euclidean) distance between codewords, ($d$). Traditionally, Hamming distance is used as the operative metric for characterizing minimum distance of a code. However, in cases when soft information is used by the decoder, Euclidean distance 	can prove to be more useful. Higher $d$ (i.e., larger distances between codewords) corresponds a more noise tolerant neural representation of space \cite{Lin1983}. In fact, ideally all errors induced by noise with amplitude less than $\floorb{\frac{d-1}{2}}$ are correctable \cite{Lin1983}\cite{Sreenivasan2011}. We computed $d$ as a function of rate, $r$, for different grid scales (Fig. \ref{fig:d_min_vs_rate}). For each configuration there is a trade-off between $d$ and $r$. Since rank tends to increase and saturate with rate, this is also a tradeoff between $d$ and $R$. When the rate is low, a low resolution of location is targeted: $d$ is larger, so more eroneous neurons may be corrected. For a fixed $r$, a lower rank yields a larger minimum distance, and lower rates give more desirable tradeoffs between rank and $d$. Note that the code with non-uniformly allocated grid cells has significantly smaller $d$ for a fixed $r$. Thus for a fixed $r$ (i.e., for codes of the same rate), the code with uniformly allocated grid cells should exhibit measurably better de-noising performance. We confirm this prediction by simulating the de-noising process and collecting statistics presented in Figures \ref{fig:PatternErrorRate} through \ref{fig:MSE}. Surprisingly, for a uniform allocation of grid cells to modules, increases in $\mu_p$ appear to effect small decreases in $d$, while when grid cells are allocated to modules non-uniformly, increases in $\mu_p$ produce small but detectable increases in $d$. Non-uniformity appears to shift the curve down and to the left. This is no surprise since non-uniform allocation decreases the total number of grid cells. Increasing $\mu_p$ results in marginal decreases in $d$, though $d$ is less sensitive to increases in phase redundancy than decreases in population size (i.e., codeword length). \par
For environments of a fixed size, $x_{\text{max}}^2 \text{cm}^2$, and a hybrid code with $N$ neurons, varying code rates implies quantizations of space with varying unit width ($\Delta L = \frac{x_{\text{max}}}{\sqrt{C}}$). Since rate, $r = \frac{C}{N}$, $\Delta L = \frac{x_{\text{max}}}{\sqrt{Nr}}$. Thus the spatial sampling period, $\Delta L$ is inversely proportional to $\sqrt{r}$. In order to ensure we probed reasonable code rates, we estimate the typical perceivable spatial period of a rat (through its place cells) by considering its running speed (ranging from .1 to 100 $\frac{\text{cm}}{\text{s}}$), and average ISI of $150$ms \cite{gupta2012segmentation}, which bounds neural sampling periods for space, implying that $\Delta L$ should lie somewhere in $[0.15, 15]$cm. Code rates considered in this work assume $\Delta L < 15$cm. To satisfy curiosity, and probe rate dependent phenomena at even greater rates, the smallest $\Delta L$ considered is $0.0022$cm.\par
\begin{figure}[H]
\centering
	\begin{subfigure}[b]{0.32\textwidth}
	\centering
	\subcaption{}
	\includegraphics[width=4.5cm,height=6.5cm]{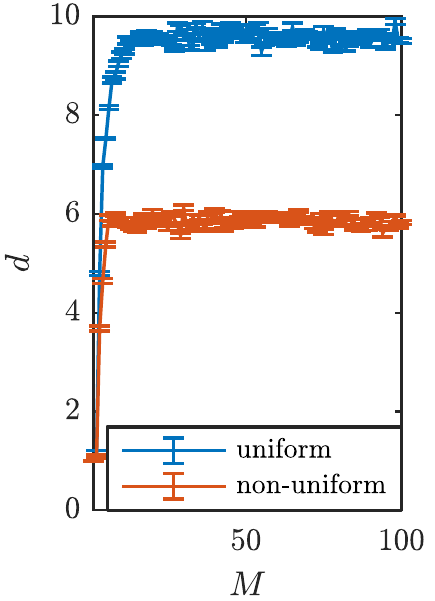}
	\label{fig:minDistVsM}
	\end{subfigure}
	\begin{subfigure}[b]{0.32\textwidth}
	\centering
	\subcaption{}
	\includegraphics[width=4.5cm,height=6.5cm]{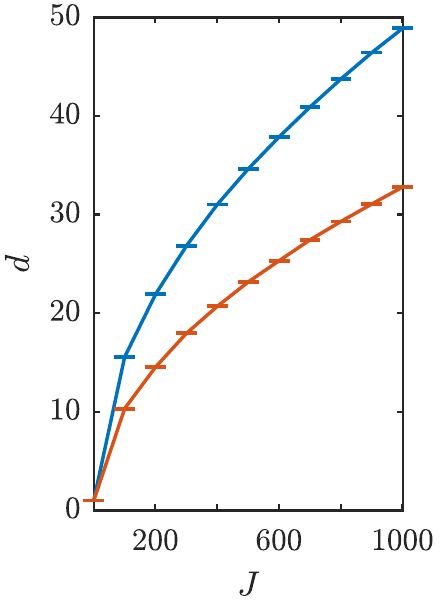}
	\label{fig:minDistVsJ}
	\end{subfigure}
	\begin{subfigure}[b]{0.32\textwidth}
	\centering
	\subcaption{}
	\includegraphics[width=4.5cm,height=6.5cm]{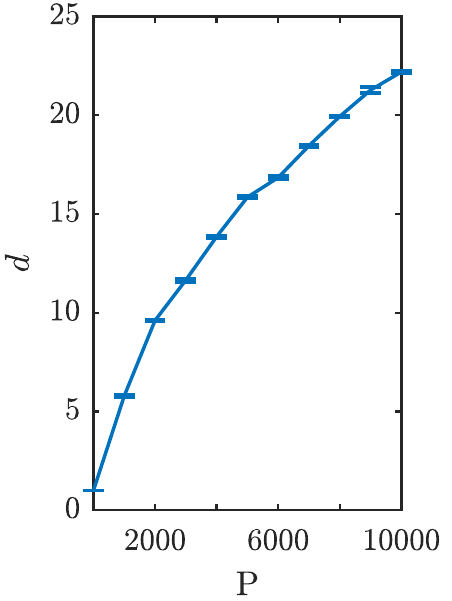}
	\label{fig:minDistVsP}
	\end{subfigure}
\caption{(a) Minimum distance, $d$, vs. $M$, with $J = 20$, $\mu_o = 6$, $\mu_p = 5$, uniform and non-uniform allocations of grid cells to modules, and no place cells. (b) $d$ vs. $J$ for uniform and non-uniform allocations of grid cells to modules, and no place cells (c) $d$ vs. $P$ for a population of place cells resembling those observed in experiment \cite{Nadel1978, MuessigHauserWillsEtAl2015, aronov2017mapping} with no grid cells. A more complete list of parameters may be found in appendix \ref{subsec:parameterChoices}}
\label{fig:minDistVsCodeSize}
\end{figure}

In order to investigate how the fundamental limits on denoisability of the code scale with the number of pattern neurons (i.e., grid and place cells), we compute $d$ as a function of $N$, independently varying $P$, $M$, $\lbrace J_i\rbrace_{i \in \lbrace 1 ,..., M\rbrace}$), fixing other paramters. As illustrated in Figure \ref{fig:minDistVsCodeSize}, minimum distance increases exponentially with increases in $N$ due to increases in number of place cells, $P$ and number of grid cells per module, $J_i$. In contrast, increases of $M$ past a critical value cease to improve minimum distance because the spatial scale at which higher order modules represent position fails to capture relevant differences in location encoded. Notably, when all other parameters are fixed, non-uniform allocations of grid cells to modules provides a code with inferior minimum distance. This is a direct result of the greater number of pattern neurons in the uniform case.\par

\subsection{Code learning results}\label{subsec:learningResults}
\begin{figure}[H]
\centering
\includegraphics[width=6.8cm,height=5cm]{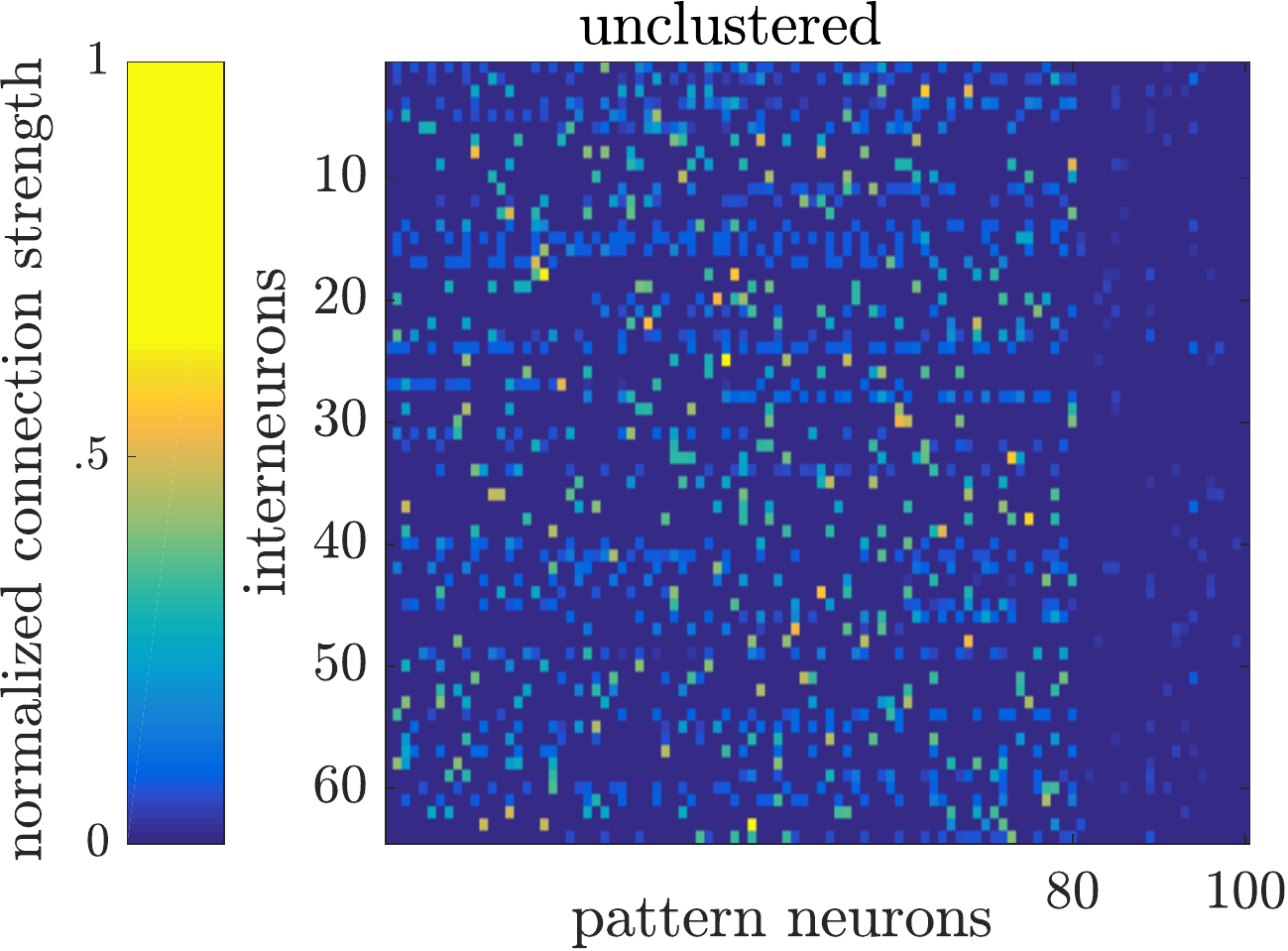}
\includegraphics[width=6.325cm,height=5cm]{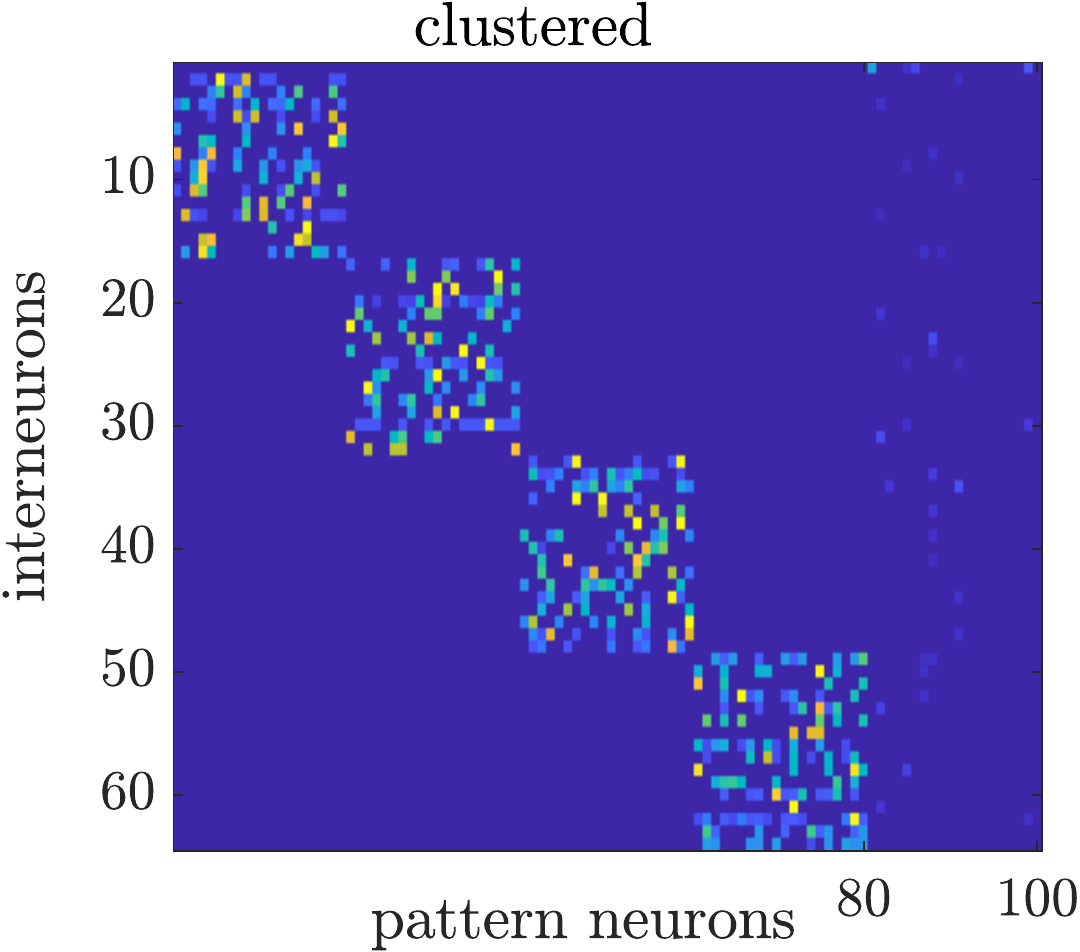}
\caption{Image of typical synaptic weight matrices learned by two de-noising networks for a hybrid code with $M = 4$, $J = 20$ and $P = 20$.}
\label{fig:weightsImages}
\end{figure}

\begin{figure}[H]
\centering
\includegraphics[width=15cm,height=5cm]{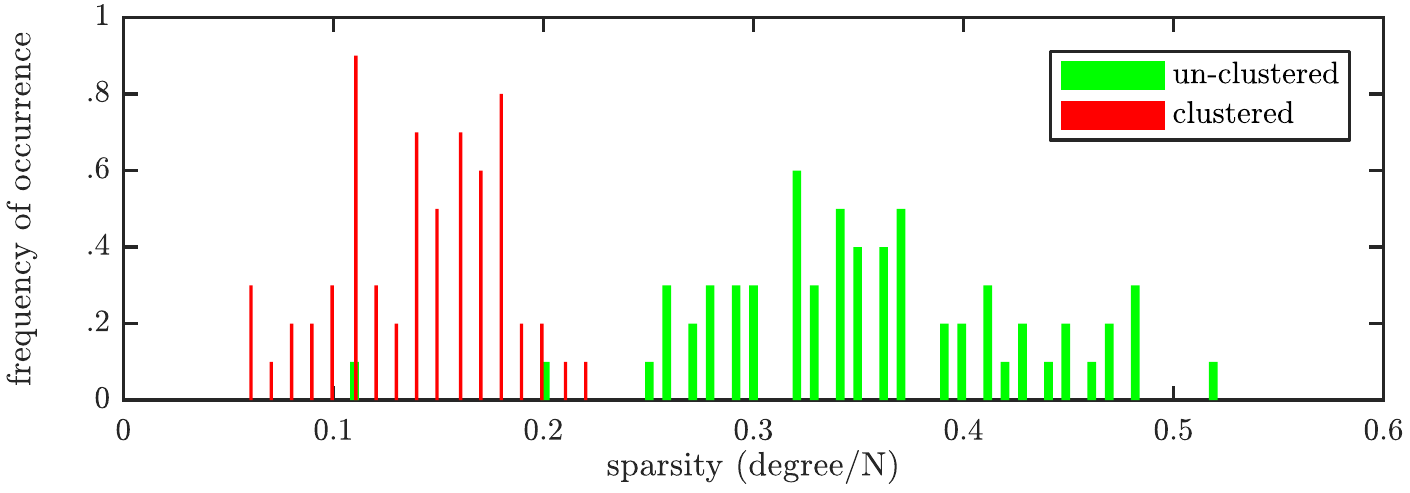}
\caption{Degree distributions of the connectivity matrices shown in Figure \ref{fig:weightsImages}}
\label{fig:degree_dist}
\end{figure}

Typical learned connectivity matrices and their associated degree distributions are found in Figures \ref{fig:weightsImages} and \ref{fig:degree_dist}. These demonstrate that for a typical hybrid code, the clustered network has a sparser connectivity, with less variability in its sparsity compared to the un-clustered network. This is because clustering enforces a tighter limit on the number of pattern neurons to which an constraint neuron may connect. We simulated an ensemble of $4$ modules of $20$ grid cells each, together with $20$ place cells, which produced the following connectivity matrices and associated degree distributions. Interestingly, in both cases, there are place cells (i.e., pattern neurons with index exceeding $80$) that are left unconnected to grid modules via constraint neurons. An illustration of the learned weights matrix corresponding to a randomly clustered de-noising network was omitted, as it is sparser, but otherwise very similar to that of the un-clustered weights image.\par

\begin{figure}[H]
\centering
\includegraphics[width=15cm,height=5cm]{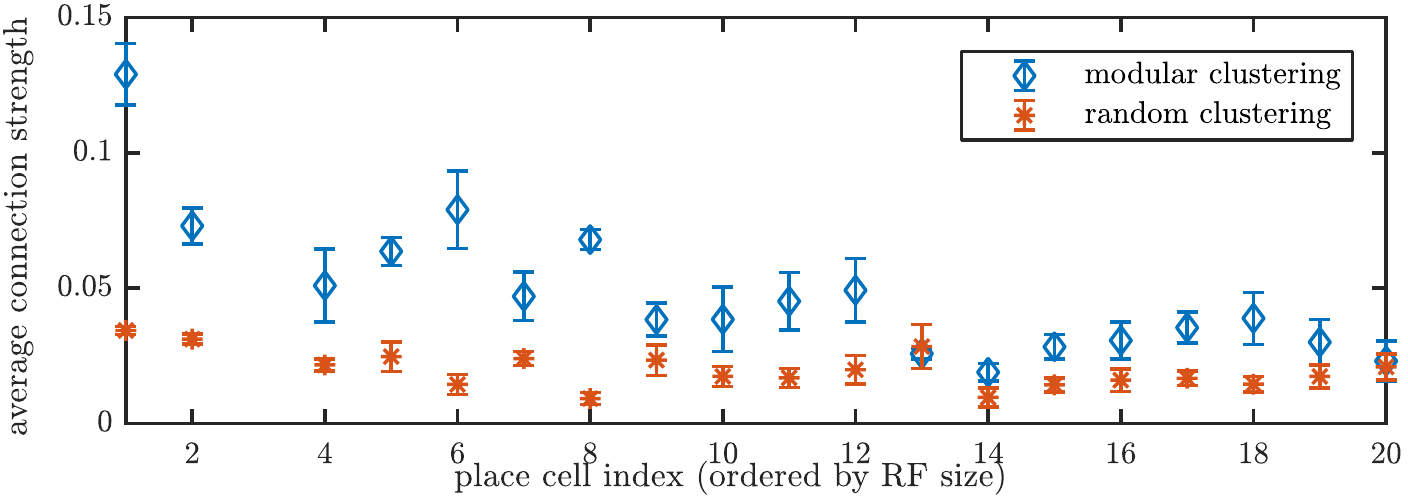}
  \caption{Average connectivities between place cells (x-axis) and grid modules for configurations with $M = 4$, $J = 20$, and $P = 20$; place cell indices are ordered from smallest to largest receptive field size; grid cell phases were uniformly distributed across the environment and grid orientations were chosen in agreement with the data presented in \cite{StensolaStensolaSolstadEtAl2012}; connection strengths depicted are averaged over $50$ networks}
  \label{fig:deliberateAndRandomClusteringConnnectivity}
\end{figure}

Figure \ref{fig:deliberateAndRandomClusteringConnnectivity} depicts the average connection strength between place cells and grid modules, where the connection strength between place cell $p$ and grid module $m$ is defined as $\frac{1}{n_i}(\sum\limits_{(i,j)} \vert w_{i,j} w_{i,p}\vert)$, where $i$ indexes constraint neurons, and $j$ indexes grid cells in module $m$. Note here that connectivity does not imply direct synaptic connection, but effective connectivity through constraint neurons. Results were obtained from configurations with $M = 4$, $J = 20$ and $P = 20$; place cells in legend are ordered by increasing receptive field size; connectivities depicted are averaged over $50$ networks. Place cells are ordered by increasing size of receptive field. This trend appears for any $\mu_p > 1$. In the modularly clustered case, average connectivity (between place cells and all grid modules) appears to decrease with increasing place cell size, as compared to a random clustering which produces nearly the same connectivity for each place cell. This phenomenon was not observed when grid phases or grid orientations were chosen randomly. It does not appear in the un-clustered configuration.\par

\subsection{De-noising and Decoding Results}\label{subsec:denoisingAndDecodingResults}
\begin{figure}[H]
\centering
\includegraphics[width=15cm, height=5cm]{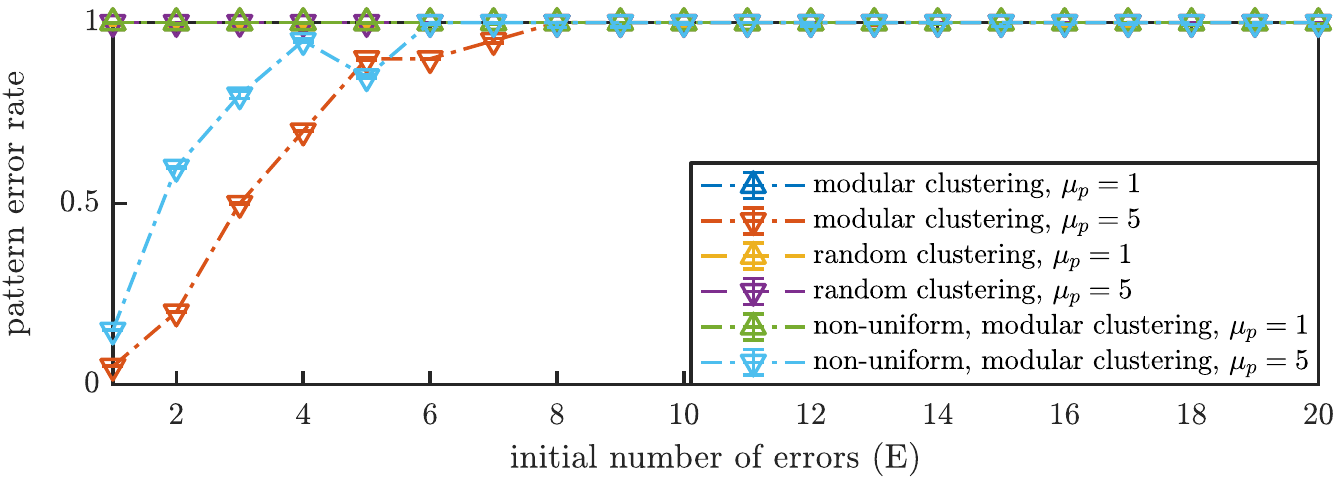}
\caption{Pattern error rate vs. initial number of errors for a clustered hybrid code with $M = 4$, $J = 20$, $P = 10$, with deliberately chosen spatial phases and orientations (i.e. so as to mirror those observed in \cite{StensolaStensolaSolstadEtAl2012}); all other pattern error rates tested (i.e., those with random redundancy parameters or those with a non-clustering denoising network) have $\text{P}_{\text{pe}} = 1$ for any initial number of errors}
\label{fig:PatternErrorRate}
\end{figure}

To measure the denoising performance of this system, we first perturb the states (i.e., firing rates) of the grid and place cells by incrementing or decrementing randomly and clipping to the boundaries of $[0, Q-1]$. A pattern error occurs if after de-noising, any entry of the de-noised pattern differs from the corresponding component of the original pattern. A symbol error occurs each time any symbol of the de-noised pattern differs from the corresponding symbol of the correct pattern. For identical populations of grid and place cells ($M = 4$, $J = 20$, and $P = 10$), in pattern error rate, the clustered network dramatically outperforms the un-clustered (when the grid cells have sufficient redundancy), and the modular clustering scheme always outperforms the random clustering scheme. By fixing the size of the populations we compare, we ensure no improvement in $d$ that results from a larger $N$. Figure \ref{fig:PatternErrorRate} depicts pattern error rate ($\text{P}_{\text{pe}}$) for a clustered hybrid code, with varying phase multiplicity and biologically observed orientation multiplicity. All other configurations (that is, those configurations that did not employ clustering in de-noising, or had randomly selected phases or randomly selected orientations) had $100$ percent pattern error rate even for a small number of initial errors. This shows that for a small number of initial errors, the full pattern of population activity corresponding to the correct location may be recovered, but that for many noise-induced errors, this is rarely possible. That only the modularly clustered de-noising network is able to achieve low $\text{P}_{\text{pe}}$ (and only when $\mu_p$ is large enough) shows that the biological organization of grid cells into discrete modules, is important for high quality self location in the presence of noise. Further, clustering is the only way to achieve such a small $\text{P}_{\text{pe}}$, since no non-clustered de-noising network consistently reduces $\text{P}_{\text{pe}}$ below $0.99$. It is no surprise that for a small number of errors, the modularly clustered de-noising mechanism achieves a better $\text{P}_{\text{pe}}$ when de-noising hybrid codes with uniform allocations of grid cells to modules (as compared to non-uniform allocations of grid cells to modules), as Figure \ref{fig:d_min_vs_rate} demonstrates that such codes tend to have a larger minimum distance at any rate probed. This result also demonstrates that whether or not grid cells are distributed uniformly to modules has a smaller impact on $\text{P}_{\text{pe}}$ than $\mu_p$.\par
\begin{figure}[H]
\centering
\includegraphics[width=15cm, height=5cm]{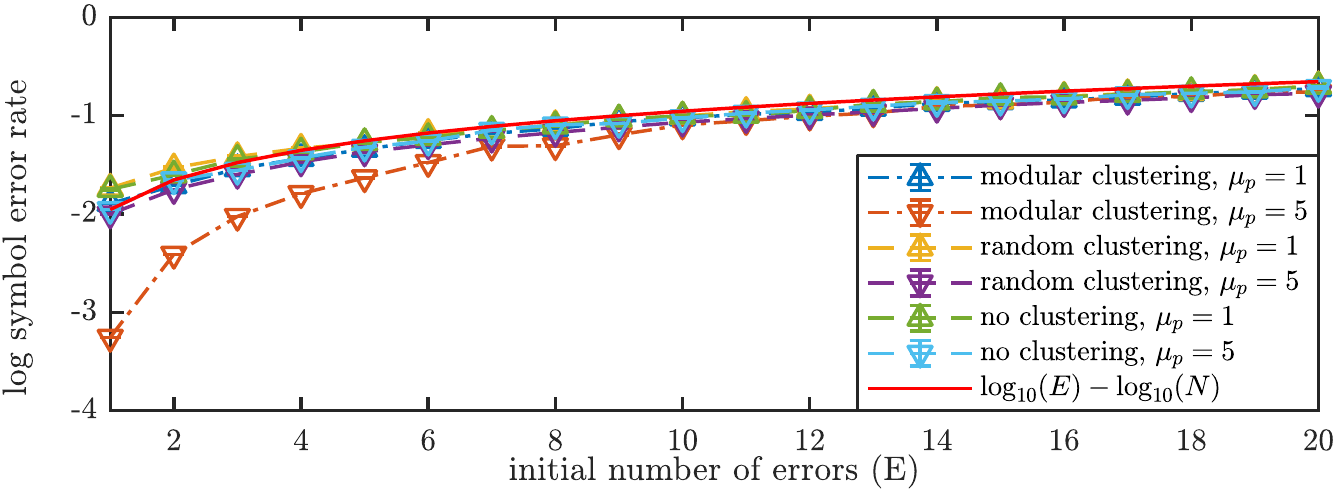}
\caption{Symbol error rate vs. initial number of errors for clustered and non-clustered hybrid codes; here, each code utilizes a uniform distribution of grid cells to modules, and deliberately chosen spatial phases and orientations (i.e. so as to mirror those observed in \cite{StensolaStensolaSolstadEtAl2012})}
\label{fig:SymbolErrorRates_nonClusteringVsClustering}
\end{figure}
\begin{figure}[H]
\centering
\includegraphics[width=15cm, height=5cm]{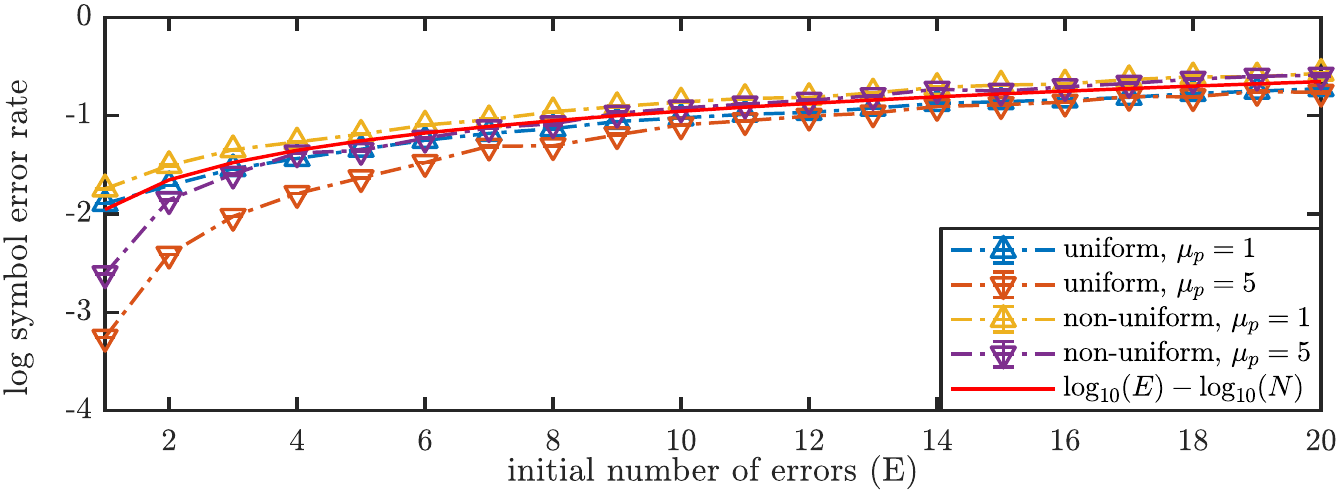}
\caption{Symbol error rate vs. initial number of errors for uniform and non-uniform clustered hybrid codes; here, each de-noising network employs the modular clustering scheme}
\label{fig:SymbolErrorRates_nonUnifVsUnif}
\end{figure}

Figure \ref{fig:SymbolErrorRates_nonClusteringVsClustering} shows symbol error rates of hybrid codes for several configurations with deliberately chosen grid cell phases and orientations (i.e. so as to mirror those observed in \cite{StensolaStensolaSolstadEtAl2012}. This demonstrates that generally, clustered de-noising networks do not offer improved symbol error rate, $\text{P}_{\text{se}}$, compared to their un-clustered counterparts. However, for a small initial number of errors, when the grid cells exhibit sufficient redundancy in their phases, a randomly clustered de-noising network is only outperformed by a modularly clustered network. Figure \ref{fig:SymbolErrorRates_nonUnifVsUnif} shows $\text{P}_{\text{se}}$ for a hybrid code with deliberately chosen phases and orientations, de-noised by a modularly clustered network. Consistent with the observations on pattern error rate, hybrid codes with grid cells uniformly allocated to modules achieve better $\text{P}_{\text{se}}$. This may result directly from the fact that $d$ is larger for such codes. On the other hand, it is possible that the smaller population sizes resulting from a non-uniform allocation (thus allowing fewer opportunities for grid cell redundancy) are behind both the improvement in denoising performance and the larger minimum distance. Plotted in both Figures \ref{fig:SymbolErrorRates_nonClusteringVsClustering} and \ref{fig:SymbolErrorRates_nonUnifVsUnif} is a red, solid curve, $\log_{10}(\frac{\text{initial number of errors}}{N})$. This curve is a threshold between regions of desirable and unacceptable $P_{\text{se}}$ (i.e., $\log_{10}(P_{\text{se}})$ for a network that performs no de-noising). To see this, consider a de-noising network that does not change the initial number of errors, $E$. For this network, $P_{\text{se}} = \frac{E}{N}$, so $\log_{10}(P_{\text{se}}) = \log_{10}(E) - \log_{10}(N)$. Surprisingly, Figure \ref{fig:SymbolErrorRates_nonClusteringVsClustering} shows that for a small initial number of errors, configurations with $\mu_p = 1$ have $\log_{10}(P_{\text{se}})$ above this threshold, that is, they increase the number of symbol errors! Figure \ref{fig:SymbolErrorRates_nonUnifVsUnif} shows that this observation is independent of uniformity of the distribution of grid cells to modules.\par

\begin{figure}[H]
\centering
\includegraphics[width=15cm, height=7.5cm]{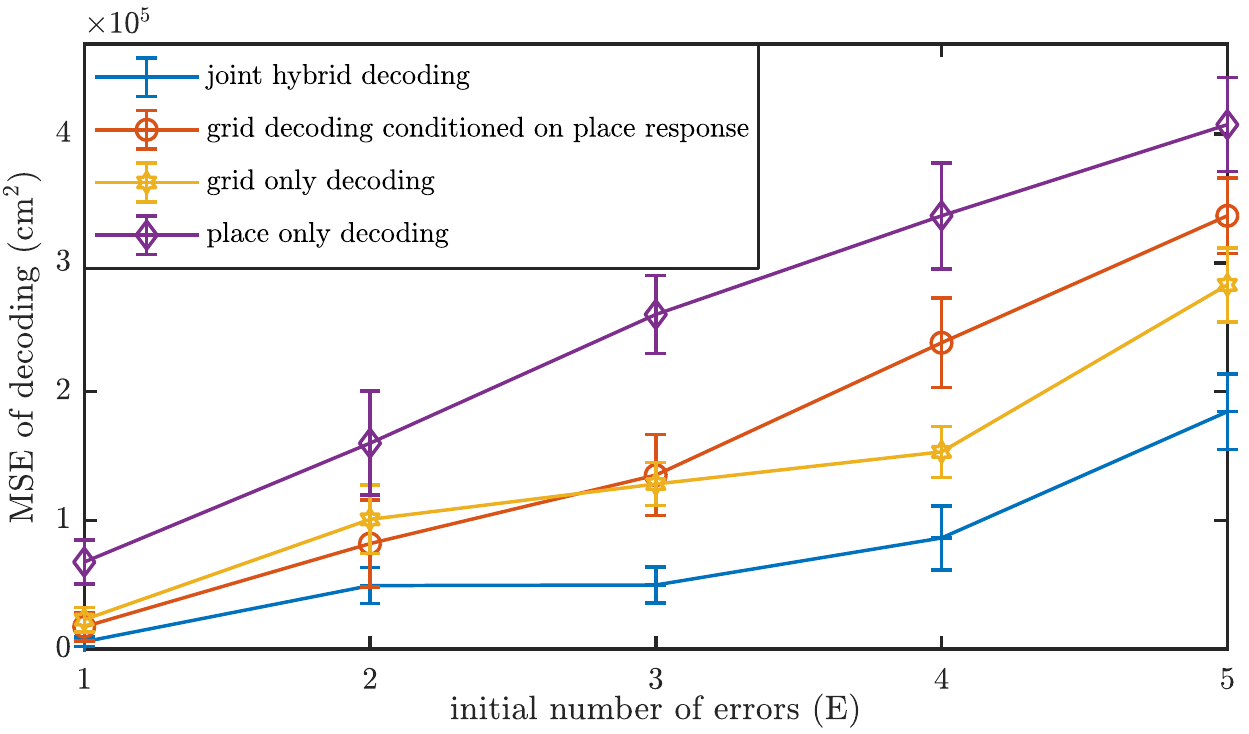}
\caption{MSE of decoding after de-noising for a hybrid code with $M = 4$, $J = 20$, $P = 10$, and $\mu_p = 5$, and deliberately chosen grid cell parameters (i.e. so as to mirror those observed in \cite{StensolaStensolaSolstadEtAl2012})}
\label{fig:MSEdecodingComparisons}
\end{figure}
\begin{figure}[H]
\centering
\includegraphics[width=15cm, height=10cm]{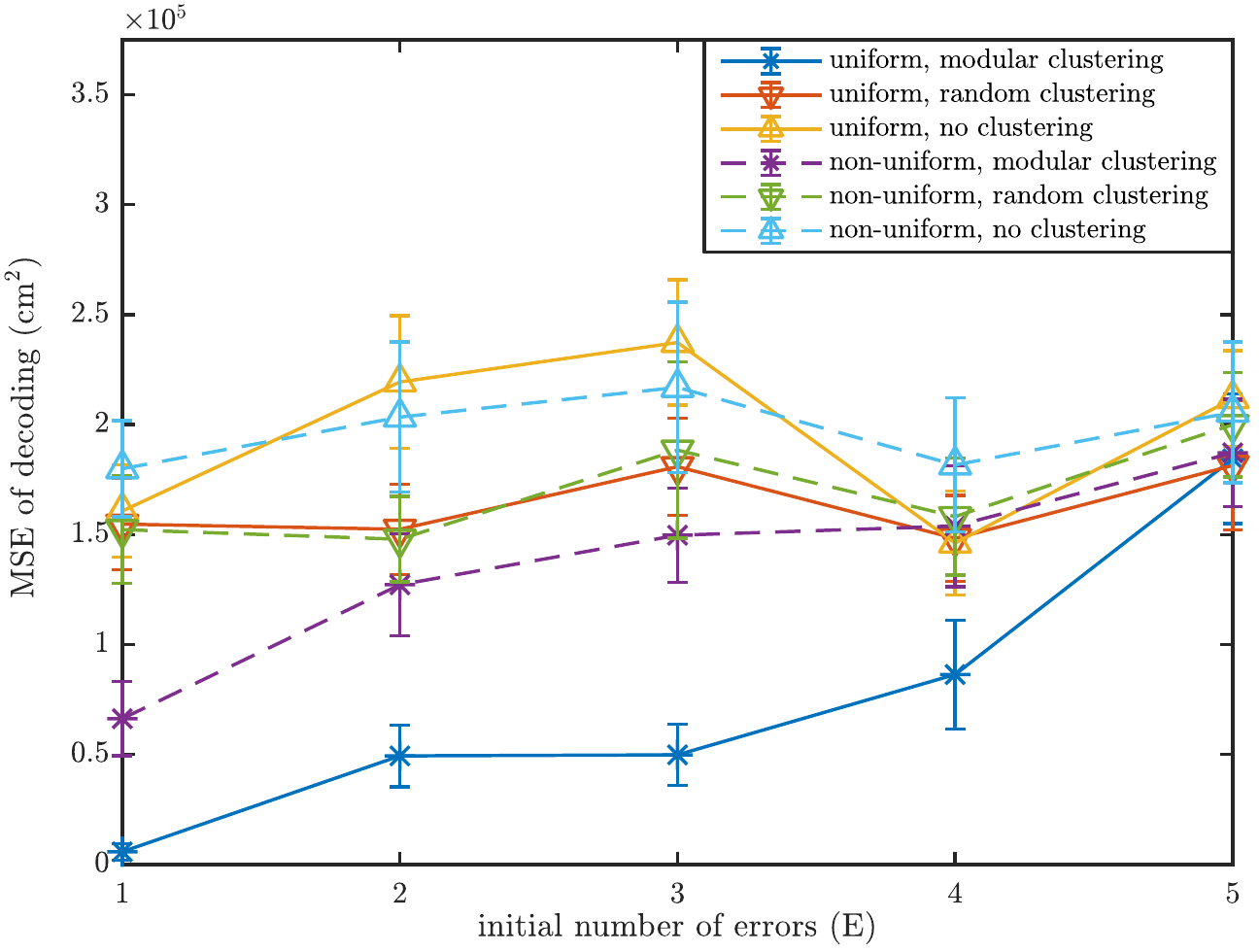}
\caption{MSE of ideal decoding after neural de-noising ($\text{cm}^2$) vs. initial number of errors for a hybrid code with $\mu_p = 5$ and $\mu_o = 6$.}
\label{fig:MSE}
\end{figure}

Figure \ref{fig:MSEdecodingComparisons} shows MSE of different decoding processes after de-noising for a Hybrid code with $M = 4$, $J = 20$, $P = 10$, and $\mu_p = 5$, for deliberately chosen grid cell parameters (i.e. so as to mirror those observed in \cite{StensolaStensolaSolstadEtAl2012}). This plot demonstrates that an ideal observer decoder which considers information from all cells outperforms all others for any initial number of errors. This disparity may, in part, be accounted for by the difference between the number of grid cells and the number of place cells. Figure \ref{fig:MSE} shows MSE of joint hybrid decoding after de-noising for a hybrid code with $\mu_p = 5$, for the configuration that achieved the best error correction performance in both $\text{P}_{\text{pe}}$ and $\text{P}_{\text{se}}$. This plot demonstrates that the code with grid cells distributed to modules uniformly, with a modularly clustered de-noising network achieves the best decoding performance, outperforming its non-uniformly distributed analogue. Since the code with a non-uniform allocation of grid cells to modules had a smaller minimum distance (compared to the same code with a uniform allocation of grid cells to modules), this result confirms our earlier hypothesis that codes with uniform allocations of grid cells across modules may be de-noised more effectively. This is no surprise since in section \ref{subsec:codingResultsSection}, we demonstrated that these codes achieve larger minimum distance as a result of the inclusion of a larger total number of grid cells. Further, this demonstrates (in a natural metric of the stimulus space) that in the most redundant hybrid code considered, a modularly clustered de-noising network is far superior to a randomly clustered or un-clustered one.  Interestingly, for a small number of initially eroneous pattern neurons, the loss (in MSE) due to a lack of modular clustering is much greater than the loss due to non-uniformity, which was also observed in Figure \ref{fig:PatternErrorRate}.\par

\section{Discussion}
\label{sec:discussion}
We demonstrated that both dense and sparse hybrid codes may be constructed by proper choice of grid and place cell parameters. We also showed that in the presence of neural noise, the activity of only those configurations with sufficient redundancy in the grid cell component of the code may be consistently de-noised. It is somewhat counterintuitive that choosing $\mu_p > 1$ produces a more noise resilient code (as shown in the de-noising performance results) as codes with uniformly allocated grid cells and largest $d$ are those with unique spatial phases (i.e., $\mu_p = 1$) (Figure \ref{fig:d_min_vs_rate}). This result is biologically counterintuitive as in \cite{Hafting2005} it is noted that the distribution of grid cell phases observed in experiment did not deviate significantly from uniformity. However, in a recently published M.S. thesis, \cite{Wennberg2015}, it is revealed that the distribution of spatial phase offsets of grid cells may be significantly non-uniform. The dataset from which this conclusion is drawn is identical to that obtained from rat 14147 in \cite{StensolaStensolaSolstadEtAl2012}. Our results imply that this observed non-uniformity in distribution of grid cell phases provides value in de-nosiability.\par
Our results reveal another suprise in Figure \ref{fig:minDistVsCodeSize}, in which the gap in minimum distance achieved between codes with uniformly and non-uniformly allocated grid cells is shown to be quite large and grow dramatically with increases $J$. Additionally, this observation demonstrates that the hybrid code for space may trade off  improvements in de-noising performance (in $d$) for efficiency of encoding ($r$) by distributing grid cells to modules non-uniformly \cite{MosheiffAgmonMorielEtAl2017}.\par
Hybrid codes of widely varying $R$, $d$, and $r$ may be instantiated by choosing appropriate parameters for the populations of grid and place cells, a fact that showcases the code's adaptability. This means that the grid and place cells may be chosen for participation in neural computations, whose performance relies on assumptions other than those presented here, which insist on a low dimensional code space, and a sparse connectivity matrix. It is particularly difficult to characterize the apparent tradeoff between $r$ and $d$, presented in Figure \ref{fig:d_min_vs_rate}, as it should indicate that $\mu_p$ has little effect on $d$, a fundamental limit of the code's de-noisability. It is quite possible that the de-noising networks presented here are incapable of achieving the codes' error correction capacity in the cases considered here. Such a scenario would allow for other properties endowed by a larger $\mu_p$ to effect the stark differences apparent de-noising efficacy of networks across codes with different values of $\mu_p$. Furthermore, this explanation seems likely, as traditional coding theory suggests that the maximum number of correctable errors in a linear block code (as a function of $d$) can be computed as $t = \floorb{\frac{d - 1}{2}}$\cite{Lin1983}. For example, in the strongest code (as measured by largest value of $d$ in Figure \ref{fig:d_min_vs_rate}) achieves $d \approx 8$ when $r$ is large, so $t \approx 3$. Figure \ref{fig:PatternErrorRate} corroborates this in demonstrating that pattern error rate saturates at $1$ for approxmiately $4$ errant pattern neurons.\par
We demonstrate that the chosen de-noising network architecture performs quite well for hybrid codes that fit its assumptions regarding rank, and poorly for those that do not. Additionally, we assessed average connectivity between place cells of varying receptive field sizes and modules of grid cells by analyzing the learned connectivity matrix. This analysis demonstrates that our model place cells of smaller receptive field size are more strongly connected to grid modules, and that they are most strongly connected to grid modules of the smallest scale. Moreover, this result presents a physiologically testable hypothesis. While difficult, two photon microscopy has been successfully employed to accurately image the microscopic structure of nervous tissue \cite{SvobodaYasuda2006}. One way to estimate connection strength between real neurons is to count the number of boutons expressed on the pre-synaptic neurons, assuming that weight should be proportional to this number, though there are may be simpler ways to estimate connection strength \cite{Bi10464}. Thus, if groups of place cells connected via constraint neurons to several distinct grid modules may be identified, this theoretical prediction - that connectivity between the hippocampus and MEC will decrease downward along the dorsoventral axis - can be confirmed or refuted. Another interesting experiment is made possible by recent advances in optogenetics, which enable single cell resolution of network activity for a population of inoculated cells (e.g. a collection of grid cells, as in \cite{Sun28072015}). While technically challenging due to the physical separation of each population in the brain, it should be possible to image simultaneous activity of grid and place cells at high temporal precision \cite{Grewe2010}. From these measurements, for a set of quantized locations, simultaneous firing rates may be estimated \cite{Theis2016}. Then, the rank, rate, and minimum distance of this empirical codebook may be computed to offer insight about limits of noise tolerance of real spatial navigation circuitry. Of particular interest is discovering the extent to which neural noise transiently varies such attributes for grid and place cells in real brains, and how these coding theoretic properties adapt (if at all) to changes in speed, context, and other variables.\par
In Figures \ref{fig:PatternErrorRate},  \ref{fig:SymbolErrorRates_nonUnifVsUnif}, \ref{fig:MSEdecodingComparisons}, and \ref{fig:MSE}, we demonstrate the differences in performance of each network structure, and of the various decoding algorithms. The universal improvements from place only decoding to joint hybrid decoding show that highly accurate position estimation can be significantly more difficult without both populations of cells. The discrepancy between `grid only decoding' and `grid decoding conditioned on place response' shows that even utilizing place cell information indirectly (by eliminating places deemed impossible given the state of the place cell population) yields a sizable improvement in decoding accuracy when there are many place cells, or when place cells are less noisy than grid cells. That the modular clustering network universally outperforms the corresponding randomly clustered network implies that the physiological organization of grid cells by their scale may provide a computational advantage in de-noising and decoding. This may be because the un-clustered network is essentially a randomly clustered network that does not take advantage of synergistic cluster computing. In any cluster, both grid cells and place cells are able to correct each others' errant activity. However, under modular clustering, in order for a grid cell in module $i$ to correct the activity of a grid cell in a different module $j$, the activity of each neuron in module $i$ must be correct so that the activity of place cells (connected to both modules $i$ and $j$) will contradict and correct the erroneous activity. \par
It should be noted here that the de-noising constraint neurons are a hypothetical construct and need not reside in the hippocampus or MEC in order to provide the previously described computations. Our conception of these constraint nodes is as single units. However, these may represent larger networks of neurons performing the identical computations. Furthermore, this work is not intended to convince readers of the necessity or existence of these cells, only to demonstrate tangible coding theoretic advantages conferred by constraint neuron moderated communication between grid and place cells. Additionally, some models of development of the grid and place cell networks demonstrate dependence between properties of each populations' apparent receptive fields that our model is unable to capture \cite{Monaco2011}. Thus, coding theoretic results presented here are confined to consideration of a more static code than what is often observed in recordings of real neuronal populations. While our model is limited in the sense that neurons are defined functionally (in contrast with biophysical models where behavior emerges from the time evolution of the model's physics), the learning algorithms considered are analogous to a Hebbian plasticity and operations required for de-noising can be feasibly implemented by networks of real neurons (if not by single units). Hence, the results discussed here have potential implications about neural codes for other continuously valued stimuli (e.g. pitch of an auditory signal, another variable encoded in the mammalian hippocampus \cite{aronov2017mapping}). \par
Further development along these threads of investigation of neural codes for space include studying coding theoretic properties of more complete navigational codes including head direction cells, boundary vector cells, and time cells \cite{Lever2009, salz2016time, Taube1990}. It would be most interesting to probe coding and information theoretic properties of place cells that encode 3D space as demonstrated to reside in the bat hippocampus \cite{Yartsev2013a}. Even with these classes of neuron, the hybrid code might be unable to encode and de-noise path information, unless there is a higher level structure that provides the encoded activity in the correct sequence. One strong candidate solution for this is to include so called hippocampal time cells. Just as place cells code for distinct locations on paths through space, time cells encode ordered moments in a temporally ordered sequence of events, precisely the information, which, when coupled with location, allows for the encoding of paths \cite{MacDonaldLepageEdenEtAl2011}.\par

\bibliographystyle{IEEEtran}
\bibliography{HybridCode}

\section{Appendix}
\subsection{Subspace learning}\label{App:learningAppendix}
In \cite{Oja1985}, the authors propose an algorithm that is capable of computing a basis for the null space of a random matrix, $A$, which is assumed to be the expected value of sample matrices, $A_t$. The update rule for the matrix whose columns are the resulting basis vectors is
\begin{equation}
\label{eq:LearningRuleMatrixForm_1}
\tilde{W}_t = W_{t -1} + A_{t-1} W_{t-1} \alphav_{t-1}
\end{equation}
\begin{equation}
\label{eq:LearningRuleMatrixForm_2}
W_t = \tilde{W}_t R^{-1}_t ,
\end{equation}
where $\alphav_t$ is a diagonal (and compatible) matrix of gain factors.
As in \cite{Oja1985}, equations \ref{eq:LearningRuleMatrixForm_1} and \ref{eq:LearningRuleMatrixForm_2} may be re-written as operations on column vectors, $\wv_t$.
\begin{equation}
\label{eq:LearningRuleColumnVectorForm_1}
\tilde{\wv}_t = \wv_{t-1} + \alpha_{t-1} A_{t-1} \wv_{t-1}
\end{equation}
\begin{equation}
\label{eq:LearningRuleColumnVectorForm_2}
\wv_t = \frac{\tilde{\wv}_t}{\Vert \tilde{\wv}_t \Vert},
\end{equation}
in which $\alpha_t$ is the gain factor corresponding to the current column. This number may be equivalently understood as a learning rate. Indeed in \cite{Xu1991}, the authors show that for appropriate choices of $A_t$, the update rule is a form of anti-Hebbian learning. In \cite{Oja1985} the authors prove convergence of this algorithm to the eigenvectors of $A$ corresponding to the largest eigenvalues. Further, when $A_t$ is replaced by $-A_t$, $\wv_t$ converges to the eigenvectors of $A$ corresponding to the smallest eigenvalues. In \cite{Oja1985}, it is demonstrated that by combining equations \ref{eq:LearningRuleColumnVectorForm_1} and \ref{eq:LearningRuleColumnVectorForm_2}, expanding as a power series in $\alpha_t$, and ignoring second (and higher) order terms, we arrive at
\begin{equation}
\label{eq:LearningRuleColumnVectorForm_3}
\wv_t = \wv_{t-1} + \alpha_{t-1}(A_{t-1}\wv_{t-1} - \frac{\wv^{T}_t A_{t-1} \wv_{t-1}} {\wv^{T}_{t-1} \wv_{t-1}} \wv_{t-1}).
\end{equation}
The authors of \cite{SalavatiKumarShokrollahi2014} choose $A_t = (\xv^T_t \xv_t)P_{\xv_t} = \xv_t \xv^T_t$, the product of projections onto the space spanned by $\xv_t$, and define $y_t = \xv^T_t\wv_t = \wv^T_t\xv_t$. In \cite{Oja1985}, it is mentioned that this update rule finds eigenvectors corresponding to the largest eigenvalue of $A_t$, or those corresponding to the smallest eigenvalues of $-A_t$, when this matrix is used instead. Since $A_t$ is a projection matrix, it has rank $1$. Thus it has one eigenvector with non-zero eigenvalue, $\xv_t$, and $\text{dim}(\xv)-1$ eigenvectors with eigenvalue $0$. Each of these eigenvectors, $\vv$, is guaranteed to be perpendicular to $\xv$ because $A_t \vv = 0\vv = \zerov$, that is, the $\vv$'s projection onto $\xv$ has magnitude $0$. By choosing $\xv_t \in \Cc$, with the aforementioned choice for $A_t$, this algorithm should compute vectors approximately perpendicular to the code space.\par
Now, we may rewrite equation \ref{eq:LearningRuleColumnVectorForm_3} as
\begin{align}
\wv_t &= \wv_{t-1} -\alpha_{t-1}\xv_{t-1} \xv^T_{t-1} \wv_{t-1} + \alpha_{t-1} \frac{\wv^T_{t-1} \xv_{t-1} \xv^T_{t-1} \wv_{t-1}}{\Vert \wv_{t-1} \Vert^2} \wv_{t-1}\nonumber\\
 &= \wv_{t-1} - \alpha_{t-1} y_{t-1} \xv_{t-1} + \alpha_{t-1} \frac{y^2_{t-1}}{\Vert \wv_{t-1} \Vert^2} \wv_{t-1}.\label{eq:LearningRuleColumnVectorForm_4}
\end{align}
To obtain a sparse basis for $\text{null}(\underbar{\Cc})$, one may add to equation \ref{eq:LearningRuleColumnVectorForm_4} a regularizing term that penalizes non-sparse solutions. In particular, using $\eta \Gamma (\wv_{t-1}, \theta_{t-1})$, as considered in \cite{SalavatiKumarShokrollahi2014}, to arrive at
\begin{equation}
\label{eq:LearningRuleColumnVectorForm_6}
\wv_t = \wv_{t-1} - \alpha_{t-1} (y_{t-1}(\xv_{t-1} - \frac{y_{t-1}\wv_{t-1}}{\Vert \wv_{t-1}\Vert^2})) - \alpha_{t-1} \eta \Gamma(\wv_{t-1}, \theta_{t-1}).
\end{equation}\par

\subsection{Structure of the performance testing simulations}
In order to evaluate the performance of the de-noising mechanisms proposed here, we first generate codes from the parameters considered in appendix \ref{subsec:parameterChoices}. Then algorithm \ref{alg:algorithm_1} is applied to the chosen de-noising network. After learning is complete, in sequence, $C$ randomly chosen codewords are corrupted and presented to the network to de-noise using algorithms \ref{alg:modular_recall} and \ref{alg:sequential_denoising}. After the de-noising process is complete, the de-noised pattern is assessed and performance is computed incrementally.

\subsection{Choices of parameters}
\label{subsec:parameterChoices}
In learning, normalized weights are initialized randomly with degree $\lceil 4 \ \log_{\text{e}}(n)\rceil$, where $n$ is the length of the weight vector. We used, $\theta_0 = 0.031$, $\eta = 0.075$, and $\alpha_0 = 0.95$. In de-noising, we set $\phi = 0.95$. Unless otherwise noted, dependent variables measured and computed are mean values averaged over $100$ networks. Error bars represent standard error of the mean.

Here we present a table of parameters indexed by figure in this manuscript. ``N/A'' in this appendix is taken to mean either that this parameter was varied or was not used.
\begin{center}
	\begin{tabular}{||c| c c c c c c c c c c c||}
	\hline
	Figure number & $L$ (cm) & $C$ & $M$ & $J$ & $P$ & $\lambda$ & $\lambda_1$ (cm) & $\mu_o$ & $\mu_p$ & $\epsilon$ & $\eta$\\ [0.5ex]
	\hline\hline
	$3$ & $300$ & $1000$ & $4$ & $20$ & N/A & $\sqrt(2)$ & $40$ & $6$ & N/A & N/A & N/A\\
	\hline
	$4$ & $300$ & $1000$ & N/A & N/A & N/A & N/A & $40$ & N/A & N/A & N/A & N/A\\
	\hline
	$5$ & $300$ & $1000$ & N/A & N/A & N/A & $\sqrt(2)$ & $40$ & N/A & N/A & N/A & N/A\\
	\hline
	$6$ & $300$ & $1000$ & $4$ & $20$ & $10$ & $\sqrt(2)$ & $40$ & N/A & N/A & N/A & N/A\\
	\hline
	$7$ & $300$ & $1000$ & $4$ & $20$ & $0$ & $\sqrt(2)$ & $40$ & $6$ & $5$ & N/A & N/A\\
	\hline
	$8$ & $300$ & $10^6$ & $4$ & $20$ & $20$ & $\sqrt(2)$ & $40$ & $3$ & $5$ & $C10^{-3}$ & $0.075$\\
	\hline
	$9$ & $300$ & $10^6$ & $4$ & $20$ & $20$ & $\sqrt(2)$ & $40$ & $3$ & $5$ & $C10^{-3}$ & $0.075$\\
	\hline
	$10$ & $300$ & $10^6$ & $4$ & $20$ & $20$ & $\sqrt(2)$ & $40$ & $3$ & $5$ & $C10^{-3}$ & $0.075$\\
	\hline
	$11$ & $300$ & $10^6$ & $4$ & $20$ & $10$ & $\sqrt(2)$ & $40$ & $3$ & N/A & $C10^{-3}$ & $0.075$\\
	\hline
	$12$ & $300$ & $10^6$ & $4$ & $20$ & $10$ & $\sqrt(2)$ & $40$ & $3$ & N/A & $C10^{-3}$ & $0.075$\\
	\hline
	$13$ & $300$ & $10^6$ & $4$ & $20$ & $10$ & $\sqrt(2)$ & $40$ & $3$ & N/A & $C10^{-3}$ & $0.075$\\
	\hline
	$14$ & $300$ & $10^6$ & $4$ & $20$ & $10$ & $\sqrt(2)$ & $40$ & $3$ & $5$ & $C10^{-3}$ & $0.075$\\
	\end{tabular}
\end{center}

\end{document}